\documentclass{aa}  
\usepackage{txfonts}
\usepackage{epic}
\usepackage{eepic}
\usepackage{graphicx}
\usepackage{epstopdf}
\usepackage{amsmath}
\usepackage{subfig}
\usepackage{xcolor}
\usepackage[]{natbib}
\usepackage{todonotes}
\usepackage{mathtools}
\usepackage{lscape}
\usepackage{amsmath}

\defcitealias{2011A&A...529A..75C}{CB11}
\defcitealias{2017A&A...600A..30C}{C17}

\title{Stellar limb darkening.}

\subtitle{A new MPS-ATLAS library for Kepler, TESS, CHEOPS, and PLATO\thanks{This work is not an official PLATO mission deliverable} passbands}

\titlerunning{MPS-ATLAS stellar limb darkening}
    
\author{N.~M.~Kostogryz\inst{1}\fnmsep\thanks{e-mail: kostogryz@mps.mpg.de}
        \and 
        V.~Witzke\inst{1} 
        \and
        A.~I.~Shapiro\inst{1} 
        \and
        S.~K.~Solanki\inst{1, 2}
        \and 
        P.~F.~L.~Maxted\inst{3}
        \and
        R.~L.~Kurucz\inst{4}
        \and
        L.~Gizon\inst{1, 5, 6}
        }
    
\institute{Max-Planck-Institut f\"ur Sonnensystemforschung,                                    Justus-von-Liebig-Weg 3, 37077 G\"ottingen, Germany\\ 
           \and
           School of Space Research, Kyung Hee University, Yongin, Gyeonggi, 446-701, Republic of Korea\\
           \and
           Astrophysics group, Keele University, Keele, Staffordshire ST5 5BG, UK\\
           \and
           Center for Astrophysics | Harvard \& Smithsonian , 60 Garden Street, Cambridge, MA 02138, USA  \\
           \and
           Institut f\"ur Astrophysik, Georg-August-Universit\"at G\"ottingen, Friedrich-Hund-Platz 1, 37077 G\"ottingen, Germany\\
           \and
           Center for Space Science, NYUAD Institute, New York University Abu Dhabi, PO Box 129188, Abu Dhabi, UAE\\
           }
\date{}

\begin{document}

 
  \abstract
   {The detection of the first exoplanet paved the way into the era of transit photometry space missions with a revolutionary photometric precision that aim at discovering new exoplanetary systems around different types of stars. With this high precision, it is possible to derive very accurately the radii of exoplanets which is crucial for constraining their type and composition. 
   However, it requires an accurate description of host stars, especially their center-to-limb variation of intensities (so called limb darkening) as it affects the planet-to-star radius ratio determination.}
   {We aim at improving the accuracy of limb darkening calculations for stars with a  wide range of fundamental parameters.}
   {We used the recently developed 1D \textbf{M}erged \textbf{P}arallelised \textbf{S}implified ATLAS (MPS-ATLAS) code to compute model atmosphere structures and to synthesize  stellar limb darkening on a very fine grid of stellar parameters. For the computations we utilised the most accurate information on chemical element abundances and mixing length parameters including convective overshoot. The stellar limb darkening was fitted using the two most accurate limb darkening laws: the power-2 and 4-parameters non-linear laws.}
   {We present a new extensive library of stellar model atmospheric structures, the synthesized stellar limb darkening curves, and the coefficients of parameterized limb-darkening laws on a very fine grid of stellar parameters in the Kepler, TESS, CHEOPS, and PLATO passbands. The fine grid allows overcoming the sizable errors introduced by the need to interpolate. Our computations of solar limb darkening are in a good agreement with available solar measurements at different view angles and wavelengths. Our computations of stellar limb darkening agree well with available measurements of Kepler stars. 
   A new grid of stellar model structures, limb darkening and their fitted coefficients in different broad passbands is provided in online tables (CDS).
   }
   {}

   \keywords{Stars: atmospheres  --
            Methods: numerical
               }

\maketitle


\newpage

\section{Introduction}\label{sect: introduction}
The discovery of the first exoplanet orbiting a solar-type star \citep{1995Natur.378..355M} has led to a rapid development of various new methods for planet detection and characterization. By now more than 3000 exoplanets have been discovered and confirmed (see Exoplanet Orbit Database\footnote{http://exoplanets.org} (EOD)).
One of the most efficient ways of detecting exoplanets is offered by the transit method \citep{2000ApJ...529L..45C}, which became particularly useful together with the high-precision photometry made possible by space telescopes such as CoRoT \citep{2006ESASP1306...11B}, Kepler \citep{2010Sci...327..977B}, TESS \citep{2015JATIS...1a4003R}, and CHEOPS \citep{2017SPIE10563E..1LC}. The transit photometry method aims at detecting periodic drops in the stellar flux due to the partial occultation of the stellar disk by the planet. The amplitude of the flux drop can be used to determine planet-to-star radius ratio, which is one of the most important parameters for characterizing an exoplanet. The transit method is also used to study atmospheres of exoplanets using transit spectroscopy, i.e., by simultaneously observing transits at several wavelengths. If the exoplanet has an atmosphere, the apparent size of the planet appears larger at wavelengths where stellar radiation is effectively absorbed or scattered by the planetary atmosphere  \citep[see, e.g.,][]{2000ApJ...537..916S, 2001ApJ...553.1006B}. Consequently, studying the dependence of the apparent planetary radius on wavelength allows constraining the chemical composition of the planet's atmosphere. The determination of planetary radius from transit measurements is one of the most important tasks in transit photometry and spectroscopy. Such a determination  requires accurate knowledge of center-to-limb variation of stellar intensity (hereafter, limb darkening) since it 
defines the shape of the transit light curve and affects its depth.

High-precision photometric measurements have shown that the present treatment of limb darkening is not sufficiently accurate and leads to systematic errors in the derived parameters of the exoplanets \cite[e.g.,][]{2016MNRAS.457.3573E, 2017EPSC...11..120M, 2018A&A...616A..39M}. Typically, limb darkening is represented by rather simple laws, such as a linear law \citep{1906WisGo.195...41S}, a quadratic law \citep{1950HarCi.454....1K}, a square-root law \citep{1992A&A...259..227D}, a power-2 law \citep{1997A&A...327..199H}, and a four-coefficients law \citep{2000A&A...363.1081C} so that when modeling the transit light curves limb darkening is parameterized by some set of coefficients. Ideally, these coefficients should be constrained by the stellar modelling. Consequently, many  libraries of limb darkening coefficients  covering a wide range of effective temperatures ($T_{\rm{eff}}$), surface gravity ($\log g$), and metallicities ($M/H$) have been produced \citep[e.g.,][]{2000A&A...363.1081C, 2010A&A...510A..21S, 2011A&A...529A..75C, 2015A&A...573A..90M}  using various radiative transfer codes, e.g., ATLAS \citep{1993KurCD..13.....K}, NextGen \citep{1999ApJ...512..377H}, PHOENIX \citep{2013A&A...553A...6H}, MARCS \citep{2008A&A...486..951G}, and STAGGER \citep{2013A&A...557A..26M}. However, the limb-darkening parameters diverge between different libraries and often lead to an inadequate quality of fits to the observed transit profiles  \citep{ 2013A&A...549A...9C}. First, this can be due to errors in limb-darkening coefficients introduced by interpolation from the grid of stellar parameters used in these libraries to the actual stellar fundamental parameters. Second, available stellar calculations may not treat mechanisms affecting limb darkening sufficiently accurately, e.g., convection \citep{2013A&A...554A.118P, 2017A&A...597A..94C} or magnetic activity \citep{2013A&A...549A...9C}. Therefore, limb-darkening coefficients are often left as free parameters in a least-squares fit to observed light curves \citep[e.g.,][]{2008MNRAS.386.1644S,2009A&A...506.1335C,2010A&A...522A.110C,2010A&A...520A..97G,2013A&A...549A...9C,2018A&A...616A..39M}. 
While this method usually leads to a good quality fit to observed transit profiles, it introduces additional free parameters, resulting in possible biases and degeneracies in the returned planetary radii \citep{2015MNRAS.450.1879E, 2017AJ....154..111M}. The way to reduce these biases and reliably determine planetary radii is to improve theoretical computations of stellar limb darkening.

With this work, we start our effort in producing a new library of stellar limb darkening. Namely, we employ the recently developed 1-D \textbf{M}erged \textbf{P}arallelised \textbf{S}implified ATLAS \citep[MPS-ATLAS,][]{2021A&A...653A..65W} code to create a grid of model atmospheres and the corresponding limb darkening for a wide range of stellar fundamental parameters, utilizing most accurate information on stellar abundances and mixing lengths. The code has been intensively tested against solar and stellar observations and incorporates improved molecular and atomic data. It is substantially faster in compare to other available codes, allowing us to calculate limb darkening on a very fine grid of stellar parameters and reduce potential errors in interpolation. 

This work is organized as follows. In Section~\ref{sect:model} we describe the computations of stellar model atmospheres on a grid of fundamental stellar parameters. We note that while the main goal of this paper is accurate calculations of the limb darkening, we also provide the grids of stellar atmospheric models which can be used for a large range of applications. 
The synthesis of stellar spectra and their convolution with different passbands is presented in Section~\ref{sect: spectralSynthesis}. Section~\ref{sect:stellar_ld} describes the derivation of coefficients for several limb-darkening laws. We compare our calculations to solar \citep{1994SoPh..153...91N} and stellar \citep{2018A&A...616A..39M} observations as well as to various available libraries of limb darkening in Section~\ref{sect: validation}. Finally, in Section~\ref{sect: summary} we summarize our results and briefly describe the outlook. 
\section{Model atmospheres} \label{sect:model}

\begin{figure}[h!]
    \centering
    \includegraphics[width=0.99\linewidth]{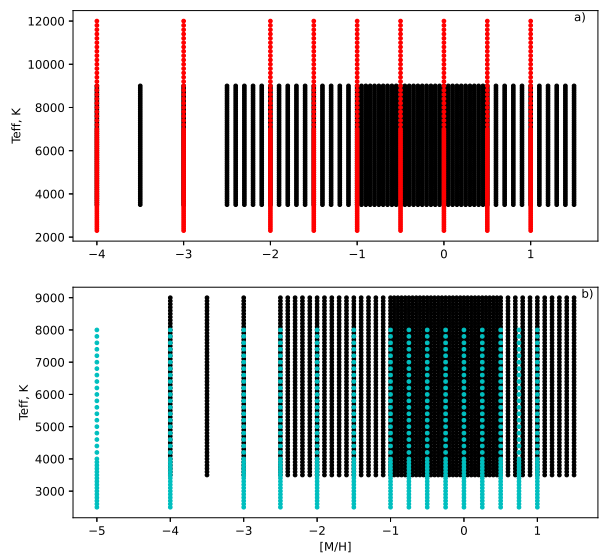}
    \caption{Representation of metallicity and effective temperature grids. Black dots show the grid from this paper. Blue dots depict the MARCS grid from \cite{2008A&A...486..951G} and red dots represents the PHOENIX grid from \cite{2013A&A...553A...6H}.} 
    \label{fig:grid_comparison}
\end{figure}

To compute the limb darkening in different broad-band passbands, we first compute 1D plane-parallel model atmospheres on a grid of fundamental stellar parameters, namely effective temperature ($T_{\rm{eff}}$), metallicity ($M/H$), and surface gravity ($\log g$) that is a lot finer than previous model atmosphere grids. 
In Table~\ref{tab:grid} we summarize the properties of our grid of stellar models.      
\begin{table}
    \centering
        \caption{Grid of stellar parameters for which we compute 1D models and limb darkening.}
    \begin{tabular}{lcc}
        Parameter & Range & Sampling\\
        \hline
        $\rm{[M/H]}$ &$-5.0$ to $-2.5$ & $0.5$ \\
        & $-2.5$ to $-1.0$ & $0.1$ \\
        & $-1.0$ to $0.5$ & $0.05$ \\
        & $0.5$ to $1.5$ & 0.1 \\
        \hline
        $T_{\rm{eff}}[\rm K]$ & $3500$ to $9000$ & 100 \\
        \hline
        $\log g$ & $3.0$ to $4.0$ & $0.5$ \\
        & $4.2$ to $4.7$ & $0.1$ \\
        & $5.0$ & \\
    \end{tabular}
    \label{tab:grid}
\end{table}
It covers the stellar parameters of most of the known exoplanet host stars (see EOD \footnote{http://exoplanets.org}) which are mainly dwarfs and sub giants of spectral types M--F. 
In Figure~\ref{fig:grid_comparison} we compare our grid of stellar models on metallicity and effective temperature to other two grids which are widely used by the community: PHOENIX \citep{2013A&A...553A...6H} and MARCS \citep{2008A&A...486..951G}. Clearly, our grid covers the largest range of metallicities with a finer sampling than the other grids. In particular, in the range of metallicities of all known stars with exoplanets ($M/H~=~[-1.0~:~0.5]$, see EOD) it has a particularly fine sampling of 0.05. In effective temperatures our grid covers a similar range and has the same temperature step as the PHOENIX models but with 100~K has only half the temperature step of the MARCS models which have a larger stepping ($200~K$) for models with $T_{\rm{eff}} > 4000~K$. In surface gravity, we have also a better sampling than the other grids, especially in the range of $\log g$ of most known stars ($\log g = [4.0 - 5.0]$). Despite the surface gravity has a small effect on stellar limb darkening, such a fine grid of model atmospheres might be useful for other purposes, e.g. for calculations of center-to-limb variation of polarization \citep[see, e.g ][]{2015A&A...575A..89K}.

For the grid of stellar parameters, we compute stellar models using the MPS-ATLAS code. This code is based on the ATLAS9 \citep{1993KurCD..13.....K} and DFSYNTHE \citep{2005MSAIS...8...34C, 2005MSAIS...8...86K} codes, which were merged into  MPS-ATLAS. MPS-ATLAS is also an extension of these codes, e.g., in its ability to treat optimum ODFs \citep{2019A&A...627A.157C}. The parallelization of the code allows performing computation much faster. Moreover, it was made more flexible and user-friendly. MPS-ATLAS includes treatment of the convection described by the mixing-length theory of turbulent transport \citep{1958ZA.....46..108B}, and parameterized by  the mixing-length parameter $\alpha$. This parameter characterizes the efficiency of energy transport in the atmosphere. It also allows the convective material to overshoot when computing the model atmosphere. 

The line blanketing, which is very important in the construction of model atmospheres, is taken into account by using the opacity distribution function (ODF) approach \citep{1984mrt..book..395C}. 
We computed ODFs using more than 100 million of atomic and molecular transitions from \cite{2005MSAIS...8...86K}. The line shape was approximated by the Voigt profile using a micro-turbulent velocity of $2~km/s$ \citep[see][]{2005MSAIS...8...86K, 2021A&A...653A..65W}.   
The continuum opacity is calculated taking into account contributions from both absorption and scattering processes. For the continuum absorption we consider free-free and bound-free transitions in $\rm{H^{-}}$, $\rm{H}$ , $\rm{He}$, $\rm{He}^-$, $\rm{C}$, $\rm{N}$, $\rm{O}$, $\rm{Ne}$, $\rm{Mg}$, $\rm{Al}$, $\rm{Si}$, $\rm{Ca}$, $\rm{Fe}$, the molecules $\rm{CH}$, $\rm{OH}$ and $\rm{NH}$, and their ions. For the scattering contribution we include electron scattering and Rayleigh scattering on $\rm{H I}$, $\rm{He I}$, and $\rm{H_2}$. A more detailed description of all physical processes and numerical schemes applied in MPS-ATLAS is given in \cite{2021A&A...653A..65W}.

\begin{figure}[!htb]
    \centering
    \includegraphics[width=0.98\linewidth]{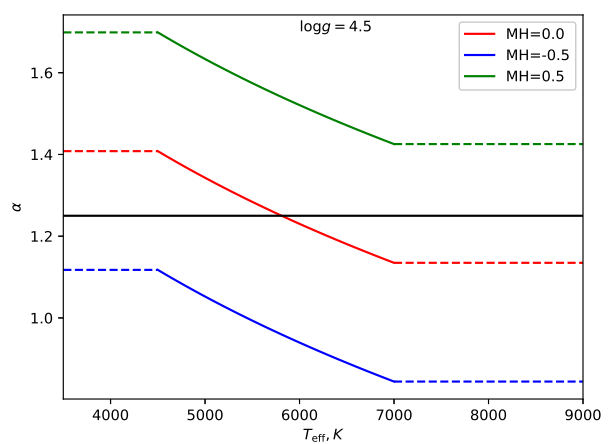}
    \caption{Mixing-length parameter dependence on stellar effective temperature for three values of metallicities ($M/H =[-0.5, 0.0, 0.5]$) and a surface gravity of $\log g = 4.5$. The solar mixing-length parameter ($\alpha_{\sun} = 1.25$) is taken from \cite{1993KurCD..13.....K}. Different colors correspond to different metallicities, as described in the legend. Dashed horizontal lines show the $\alpha$ values that are used in computations outside the region of effective temperatures for which the mixing-length behavior was derived by \cite{2018ApJ...858...28V}.}
    \label{fig:mixing_length}
\end{figure}

In addition to the grid of fundamental parameters there are two further important inputs to our simulations, namely chemical composition and mixing-length parameters. The previous grid of ATLAS9 model atmospheres by \cite{2003IAUS..210P.A20C} was computed utilising \cite{1998SSRv...85..161G} abundances and prescribing a mixing-length parameter of  $\alpha=1.25$ for all computed atmospheres.
As MPS-ATLAS is based on the ATLAS9 code we first present
the set of models with the same abundances and mixing-length parameter. Hereafter, we call this set of models  as ``set~1''. Our second set of models  (``set~2'') is based on more recent chemical abundances \citep{2009ARAA..47..481A}. For set~2 we also accounted for the dependence of the mixing-length parameter on stellar fundamental parameters ($T_{\rm{eff}}$, $M/H$, and $\log g$) using  an approximation on $\alpha/\alpha_\sun$
derived by \cite{2018ApJ...858...28V} from asteroseismic measurements of about 450 stars in the range of stellar parameters of $4500~K \le T_{\rm{eff}} \le 7000~K$, $-0.7 \le M/H \le 0.5$, and $3.3 \le \log g \le 4.5$. We scaled these dependencies to $\alpha_\sun=1.25$ \citep[as proposed by][]{1993KurCD..13.....K}.
In Figure~\ref{fig:mixing_length} we present an example of $\alpha$ dependence on stellar effective temperature for three different values of metallicity and the same surface gravity ($\log g = 4.5$). We do not extrapolate $\alpha$ outside the ranges of stellar parameters for which the relationship was derived, but take the boundary value as a constant $\alpha$ for all models outside the range (the dashed lines in Figure~\ref{fig:mixing_length}). 
In Table~\ref{tab:sets} we summarize the description of different sets of stellar model grid computations.

\begin{table}[]
    \centering
    \caption{Chemical abundances and mixing-length parameters used in our sets of stellar models}
    \begin{tabular}{ccc}
        sets & chemical & mixing-length\\ &abundances& parameter ($\alpha$) \\
        \hline
        set 1 & $1$ & $1.25$\\
        set 2 & $2$ & $0.93$ to $2.23$\tablefootmark{$\dagger$}\\
    \end{tabular}
    \tablefoot{
    \tablefoottext{$\dagger$}{The $\alpha$ values are taken from \cite{2018ApJ...858...28V} for various fundamental stellar parameters and normalized to the solar value of $\alpha_{\rm{\sun}}=1.25$}.
    \tablebib{
    (1)~\citet{1998SSRv...85..161G}; (2) \citet{2009ARAA..47..481A}.}
    }
    \label{tab:sets}
\end{table}

To compute a model atmosphere, the MPS-ATLAS code takes an initial starting model and iterates it until radiative equilibrium (RE) is satisfied. We note that in the upper convection region, MPS-ATLAS allows the model to deviate from RE due to convection and overshooting \citep[see details in][]{2021A&A...653A..65W}. The model is converged if the maximum relative temperature deviation at each point in the atmosphere is less than $10^{-5}$ or until it reaches the maximum number of iterations, which is set to 1000. The latter case happens rarely and mostly for hot metal-rich stars. For comparison, we note that ATLAS9 models by \cite{1993KurCD..13.....K} and \cite{2005MSAIS...8...34C} are computed considering only 15 iterations. For the set~1 of models, we take starting models with the closest fundamental parameters from the ATLAS9 grid. For the set~2 of models grid, the starting model is taken from the set~1 with the corresponding fundamental parameters. We note, however, that change of the starting model does not affect the final model atmosphere structure (but might strongly affect the number of iterations needed to reach the convergence). 

\begin{figure*}[!htb]
\centering 
    \includegraphics[width=0.98\linewidth]{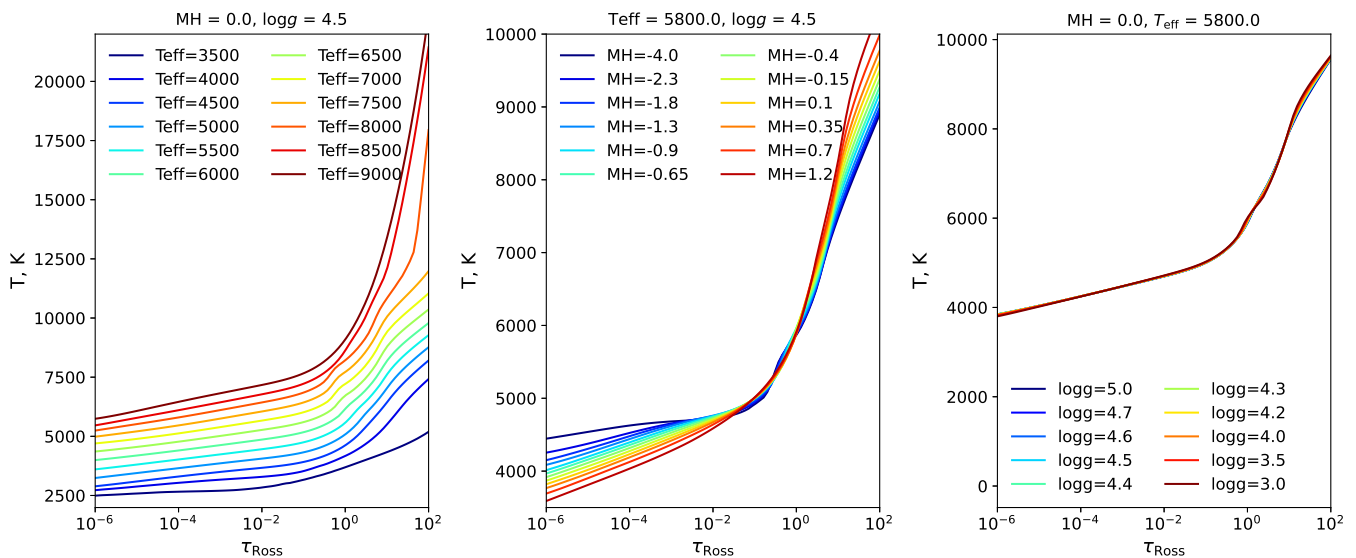}\\
    \caption{Temperature structure of several models from set~1 plotted as a function of Rosseland optical depth. Three panels show models when one of the three stellar parameters is varied and the other two are fixed. Fixed parameters are presented in the title to each panel and the varying parameter in the legend. We plot every 5th model from the effective temperature and metallicity grids.}
    \label{fig:model_atmospheres_set1}
\end{figure*}

In Figure~\ref{fig:model_atmospheres_set1} we present an example of the temperature structures computed for the set~1 of model atmospheres. To illustrate how the temperature stratification depends on different stellar parameters, i.e., effective temperature, surface gravity, and metallicity, we fix  two of them and vary the third. Figure~\ref{fig:model_atmospheres_set1} shows that increasing the $T_{\rm{eff}}$ shifts the entire structure to higher temperatures, but also produces some changes in the details of the stratification. The effect of metallicity on temperature structures is different. For a fixed effective temperature, metal-poor stars have smaller opacities and therefore require a smaller temperature gradient in order to transport the same energy through the atmosphere (middle panel in Figure~\ref{fig:model_atmospheres_set1}). Vice versa, metal-rich stars have larger opacities and thus steeper temperature gradients. 
The surface gravity does not affect the model structure computation, except for  small effects in the region where convection and overshooting play a role.
We note that  we allow the convective material to overshoot up to one pressure scale height for almost all models except of some models with low effective temperatures and low metallicities, where we suppress overshoot to avoid divergence in iterations of model atmosphere structures. The energy influx brought by the overshoot causes a decrease of the temperature gradient. It is visible as a bump at $\tau_{\rm{Ross}} \approx 1$ (where $\tau_{\rm{Ross}}$ is the Rosseland optical depth) in the atmospheric temperature structures of stars with  $T_{\rm{eff}} \gtrsim$~6000~K. For the same stars at $\tau_{\rm{Ross}} \approx 10$ another decrease of temperature gradient caused by hydrogen ionization happens and becomes prominent as a second bump in the temperature structure. 

The two sets of stellar model atmospheres are uploaded as online material of this paper (in CDS). In addition to model structures, the uploaded files also contain information about chemical composition, convection, and overshoot parameters as well as maximum temperature alteration at the last iteration (which is $10^{-5}$ in most cases but can be larger for a couple of models if 1000 iterations were not enough for convergence). In the Appendix~\ref{aa: model}, we describe an example of a model structure with all additional parameters we provide.

\section{Spectral synthesis} \label{sect: spectralSynthesis}
The spectral synthesis computations for the two sets of model atmospheres are performed using the same radiative transfer solver, turbulent velocity, and continuum opacities as used for computing atmospheric models in Sect.~2. 

We computed the limb darkening profiles for the passbands of instruments that are widely used to detect and characterize transiting exoplanets, i.e., Kepler \citep{2010Sci...327..977B}, CHEOPS \citep{2017SPIE10563E..1LC}, and TESS \citep{2015JATIS...1a4003R}. We also provide preliminary (not official)  limb darkening profiles for the upcoming PLATO mission \citep{2014ExA....38..249R}. The detectors of the telescopes are charged-coupled devices (CCDs) which count photons and do not measure energy directly. Thus, to compute the specific intensity, $I(\lambda, \mu)$, emitted from the stellar atmosphere at different wavelengths ($\lambda$) and view angles ($\mu = \cos(\theta)$, where $\theta$ is the angle between the normal to the stellar surface and the direction to the observer) we need to convert units from the energy flux to photon number flux: 

\begin{equation}
    I_{pb}(\mu) = \int_0^\infty R_{pb}(\lambda) I(\lambda, \mu) / (hc/\lambda)\ {\rm d}\lambda,
    \label{Eq: filter_convolution}
\end{equation}

\noindent where $R_{pb}$ is the response function of the instrument with a CCD detector, ``$pb$`` denotes the passband of different space missions, $h$ is the Planck constant and $c$ is the speed of light. This transformation strongly affects the spectral profile of emergent intensity in  broad-band passbands.

\begin{figure*}[!htb]
    \centering
    \includegraphics[width=0.51\linewidth]{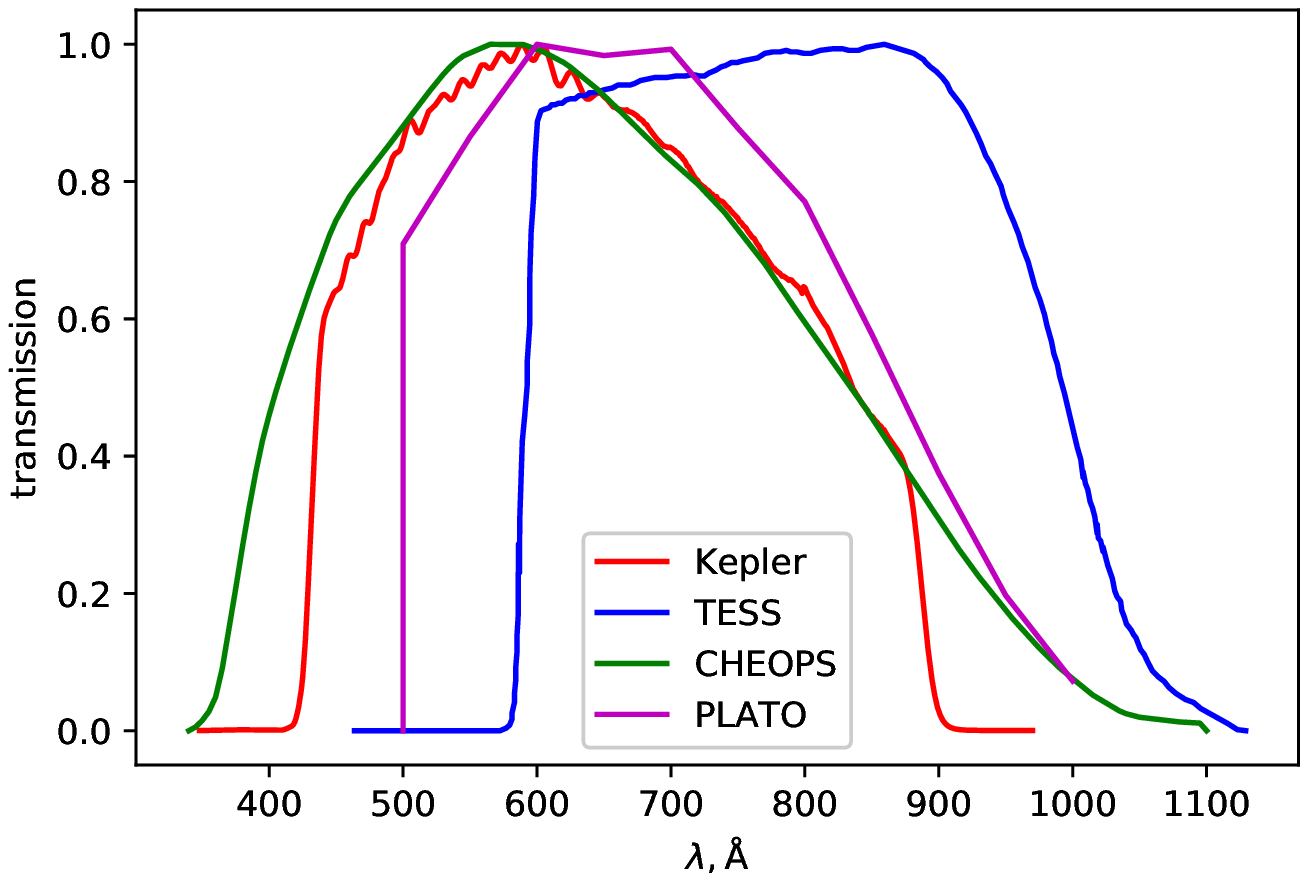}
    \includegraphics[width=0.45\linewidth]{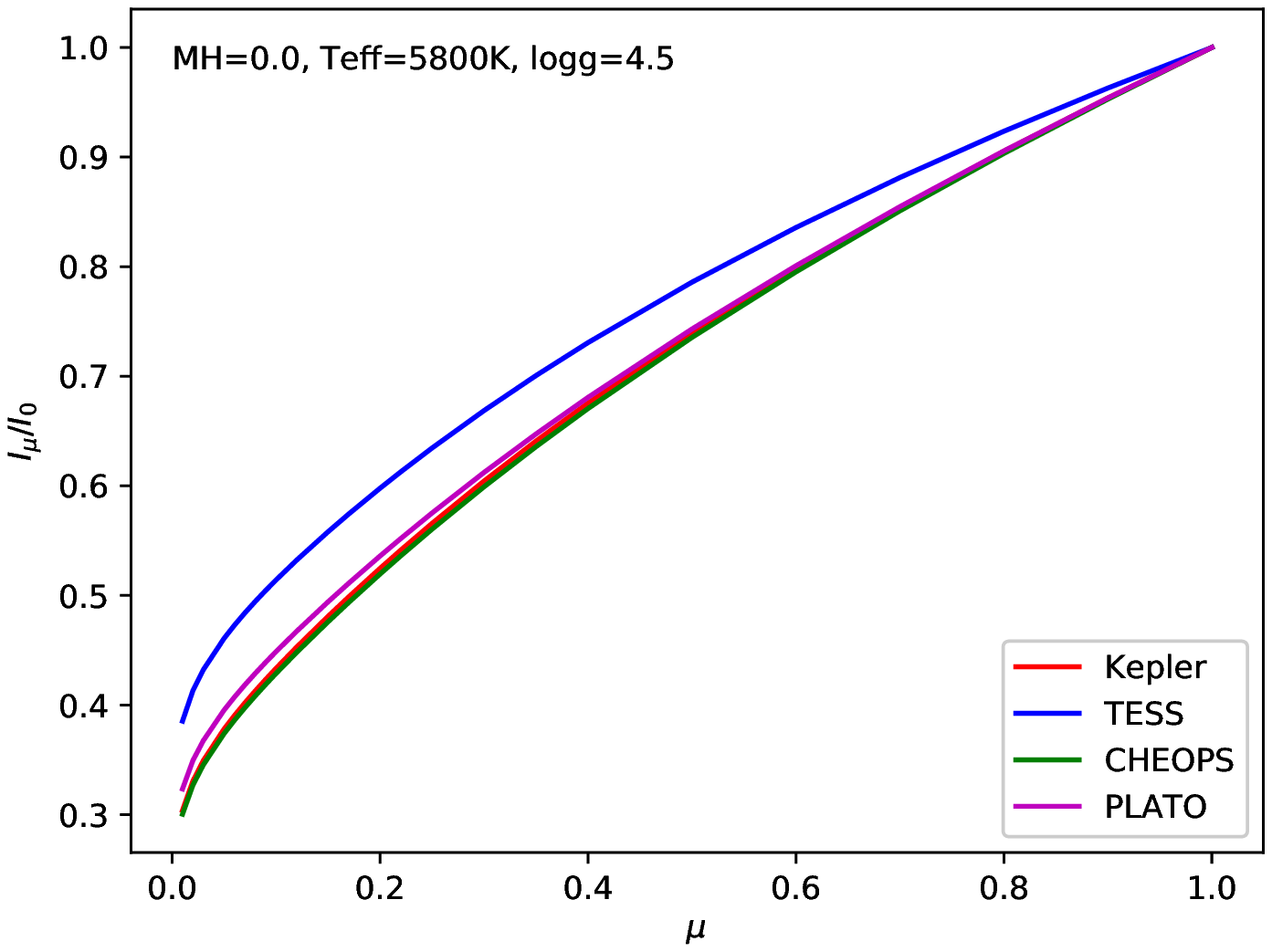}
    \caption{Normalized broad-band transmission curves of the exoplanetary space missions, Kepler, TESS, CHEOPS, and PLATO considered in this study (left panel) and the corresponding limb darkening for a star with $M/H=0.0$, $T_{\rm{eff}} = 5800$~K, and $\log g = 4.5$ (right panel). Different colors indicate different filters described in the legend. }
    \label{fig:filters}
\end{figure*}

In Figure~\ref{fig:filters} we present the transmission curves of the (broad) passbands considered in this paper together with the example of limb darkening curves in these filters for a solar analog with $M/H = 0.0$, $T_{\rm{eff}}=5800$~K, $\log g=4.5$. We note that for the PLATO passband we used the preliminary transmission curve published by \citep{2019A&A...627A..71M}, which may change in future.  Figure~\ref{fig:filters} shows that the limb darkening in Kepler, CHEOPS, and PLATO are very close to each other, with some small deviation at the limb in the PLATO passband, while limb darkening in the TESS passband deviates substantially for the same target. The deviations in the TESS passband are because it is shifted to longer wavelengths and transmits stellar radiation in the near-infrared region cutting all the radiation below 600~nm (see left panel in Fig.~\ref{fig:filters}).

We utilise the ODFs to calculate the specific intensity, $I(\lambda, \mu )$, in Eq.~\ref{Eq: filter_convolution}. While the ODFs approach has been routinely used in calculating grids of limb darkening \citep[e.g.,][]{2011A&A...529A..75C},
there have been concerns that its performance might deteriorate for cool metal-rich stars \citep{1986A&A...167..304E}.  In Appendix~\ref{aa: ODF} we introduce a modification of the ODFs setup for cool metal-rich stars which leads to a more accurate spectral synthesis. We tested limb darkening profiles calculated with the ODFs approach against those calculated utilising spectral calculations with high-spectral resolution (R=500'000) and show that our setup allows a very accurate calculations of limb darkening.

In the online tables (CDS) we provide the absolute value of intensity at the disk center and the center-to-limb intensity normalized to the disk-center intensity for 24 disk positions in different passbands for two sets of our grid. We present an example of the table in Appendix~\ref{aa:clv}.
  
\section{Parameterization of stellar limb darkening} \label{sect:stellar_ld}

Intensities, normalized to the disk center, $I_{pb}(\mu)/I_{pb}(1.0)$ are often approximated by different limb-darkening laws that are  linear or non-linear polynomial expansions in $\mu$. Different authors presented a variety of limb darkening-laws, e.g., linear \citep{1921MNRAS..81..361M}, quadratic \citep{1950HarCi.454....1K}, square-root \citep{1992A&A...259..227D}, logarithmic \citep{1970AJ.....75..175K}, power-2 \citep{1997A&A...327..199H, 2018A&A...616A..39M}, four-parameter non-linear \citep{2000A&A...363.1081C}. The four-parameters law describes the detailed shape of the limb-darkening profile for a given filter in the most accurate way \citep[e.g.,][]{2020AJ....159...75M}.  It is written as follows, 
\begin{equation}
    \frac{I_{pb}(\mu)}{I_{pb}(1.0)} = 1 - \sum_{j=1}^4 a_j (1 - \mu^{j/2}),
    \label{eq:claret_law}
\end{equation}
\noindent where $a_j$ are four coefficients of the limb-darkening law.  
Among limb-darkening laws with two coefficients, \cite{2017EPSC...11..120M} found that a power-2 law outperforms other laws, in particular, for cool stars. This law is presented as
\begin{equation}
    \frac{I_{pb}(\mu)}{I_{pb}(1.0)} = 1 - c(1-\mu^\alpha),
    \label{eq: power-2}
\end{equation}
\noindent where $c$ and $\alpha$ are the two coefficients that are a function of stellar parameters. This law matches  accurately the limb-darkening profile due to the exponent of $\mu$. 

In this paper, the center-to-limb variations of intensity convolved with the transmission profiles of different space mission instruments are fitted by both limb-darkening laws: given by Eq.~\ref{eq:claret_law} and by Eq.~\ref{eq: power-2}. We use a non-linear least squares fitting procedure from the python scipy library\footnote{scipy.optimized.curve\_fit} to fit the limb darkening curves of our grids by these two laws. The derived coefficients of the power-2 (Eq.~\ref{eq: power-2}) and the four-coefficients (Eq.~\ref{eq:claret_law}) laws are provided in online tables for the two grid sets in all passbands. An example of the table is given in Appendix~\ref{aa: LD_coefficients}. We note, however, that since limb darkening curve can not be exactly represented by parameterization, the fitting procedure might introduce biases (see Appendix~\ref{aa:h1h2_biases} for an example).
We provide a list of all online tables with titles in Table~\ref{tab:list_of_tables}.

 \begin{table}
      \caption[]{List of the provided tables which are available in CDS.}
         \label{tab:list_of_tables}
     $$ 
         \begin{array}{p{0.2\linewidth}p{0.5\linewidth}p{0.1\linewidth}}
            \hline
            \noalign{\smallskip}
            List of tables in CDS  &  Description & grid sets  \\
            \noalign{\smallskip}
            \hline
            \noalign{\smallskip}
                Table 1 & Stellar model atmosphere structures & set 1\\
                Table 2 & Stellar model atmosphere structures & set 2\\
                Table 3 & Stellar CLVs in the Kepler, TESS, CHEOPS, PLATO passbands & set 1\\
                Table 4 & Stellar CLVs in the Kepler, TESS, CHEOPS, PLATO passbands & set 2\\
                Table 5 & Stellar limb darkening coefficients in the Kepler, TESS, CHEOPS, PLATO passbands & set 1\\
                Table 6 & Stellar limb darkening coefficients in the Kepler, TESS, CHEOPS, PLATO passbands& set 2\\ 
            \noalign{\smallskip}
            \hline
         \end{array}
     $$ 
   \end{table}

\begin{figure}
    \centering
    \includegraphics[width=0.95\linewidth]{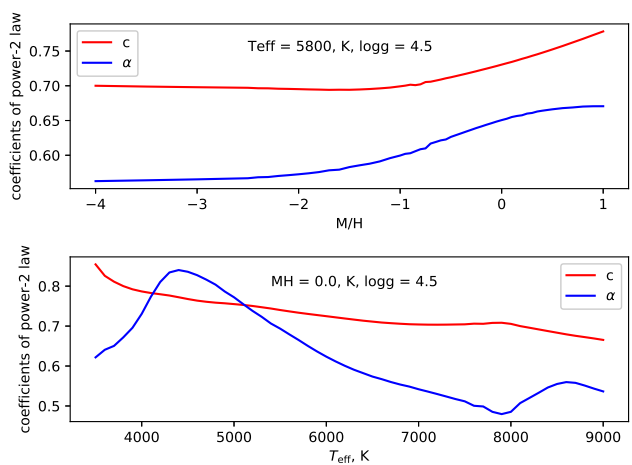}
    \caption{Limb-darkening coefficients of power-2 law on the stellar metallicity (top panel) and effective temperature (bottom panel) grids in Kepler passband. Different colors depict the coefficients ($\alpha$ and $c$). Fixed parameters are labeled on each panel.}
    \label{fig:ld_coefficients}
\end{figure}

An important feature of our calculations is that they are available on a very fine grid of fundamental parameters (in particular in metallicity, see Fig.~\ref{fig:grid_comparison}). This allows minimizing the error associated with the interpolation
of limb darkening coefficients from the grid points to specific stellar parameters (see Sect.~\ref{sect: introduction}). 
As an example in Figure~\ref{fig:ld_coefficients}, we show how the power-2 law coefficients, namely $\alpha$ and $c$ from set~1 in Kepler passband, vary with stellar metallicity and effective temperature. Note that the variation of the coefficients on stellar surface gravity is tiny, so we do not present it in Figure~\ref{fig:ld_coefficients}. One can see that the dependence of the coefficients on stellar parameters is not linear. This non-linearity causes the errors when using linear interpolation, especially when the step between the grid points is not fine enough. 

\begin{figure*}[h!]
    \centering
    \includegraphics[width=0.47\linewidth]{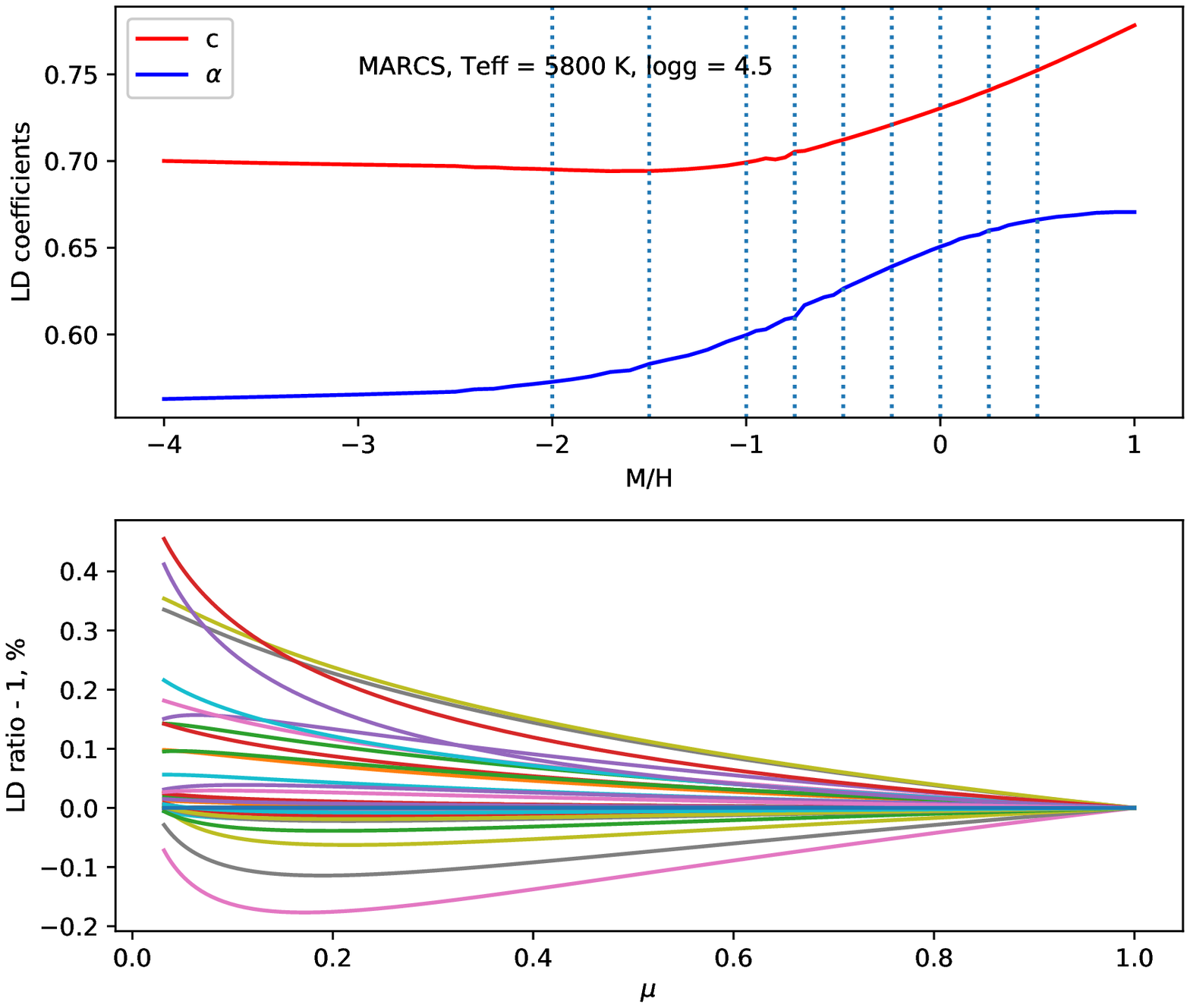}
    \includegraphics[width=0.47\linewidth]{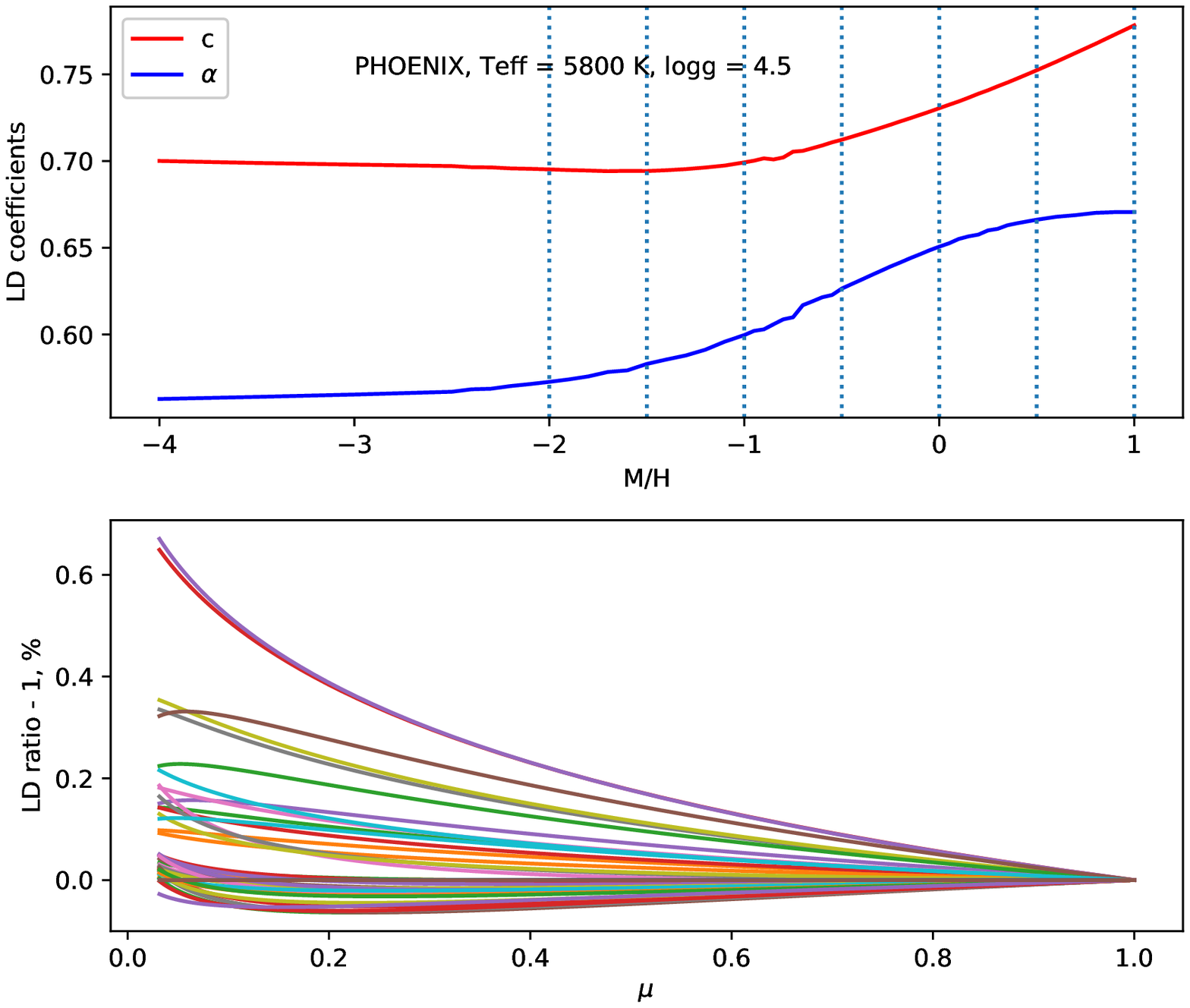}
    \caption{Top panels: Limb darkening coefficients of the power-2 law as a function of stellar metallicity in Kepler passband. Vertical dashed lines show the metallicity grid points of MARCS and PHOENIX models (left and right panels, respectively). Bottom panels: Ratios between the limb darkening curves produced from coefficients directly calculated on our fine metallicity grid and those produced from the same coefficients but selected on the MARCS and PHOENIX metallicity grid points and linearly interpolated to our fine metallicity grid (left and right panels, respectively). Ratios corresponding to different metallicity values are indicated by different colors. Coefficients and the limb darkening ratios are presented for the fixed $T_{\rm{eff}} = 5800$~K and $\log g = 4.5$.}
    \label{fig:ld_interpolation_error}
\end{figure*}

To give an example of possible values of the interpolation errors, we selected our limb-darkening coefficients from ``set~1`` on the metallicity grids of MARCS and PHOENIX models and linearly interpolated them to our metallicity grid (see top panels of Figure~\ref{fig:ld_interpolation_error}).  
In the bottom panels of Figure~\ref{fig:ld_interpolation_error} we  present the ratio of the limb darkening curves directly derived from our ``set~1`` coefficients to the limb darkening derived from the interpolated coefficients for metallicity in the range from -2.0 to 0.5 for the MARCS models and -2.0 to 1.0 for the PHOENIX models. Interestingly, Fig.~\ref{fig:ld_interpolation_error} shows that for stars with solar effective temperature the errors associated with the interpolation  from MARCS and PHOENIX grids can reach 0.4\% and 0.6\%, respectively. The interpolation errors increase towards later spectral types and decrease towards earlier spectral types (compare Fig~\ref{fig:interpolation_error_9000} to Figs.~\ref{fig: interpolation_error_3500}~and~\ref{fig:selected_interpolation_error_3500}).

\section{Validation of limb darkening computations} \label{sect: validation}
In this section, we present the validation of our computed limb darkening for the Sun and other stars by comparing the computations with solar \citep{1994SoPh..153...91N} and stellar \citep{2018A&A...616A..39M} measurements, respectively, as well as with other computations \citep[e.g., ][]{2010A&A...510A..21S, 2011A&A...529A..75C, 2017A&A...600A..30C}. 

\subsection{Solar computation-to-measurements comparison}\label{ss: validation1}
Here we compare our calculations for a star with solar fundamental parameters to available solar measurements. 
Various authors \citep[e.g.,][]{1977SoPh...52..179P, 1994SoPh..153...91N} measured the solar continuum spectrum in a broad wavelength range at different view angles. In this work, we make use of the observations by \cite{1994SoPh..153...91N} as they cover the entire wavelength range of passbands considered in our study. 
The limb-darkening computations are done for the solar parameters of $M/H=0.0$, $T_{\rm{eff}}=5777~K$, $\log g = 4.4377$ and solar abundances from \citet{2009ARAA..47..481A}, i.e., they correspond to set 2.

\begin{figure*}[!htb]
\centering 
    \includegraphics[trim=50 50 60 50,clip, width=1.0\linewidth]{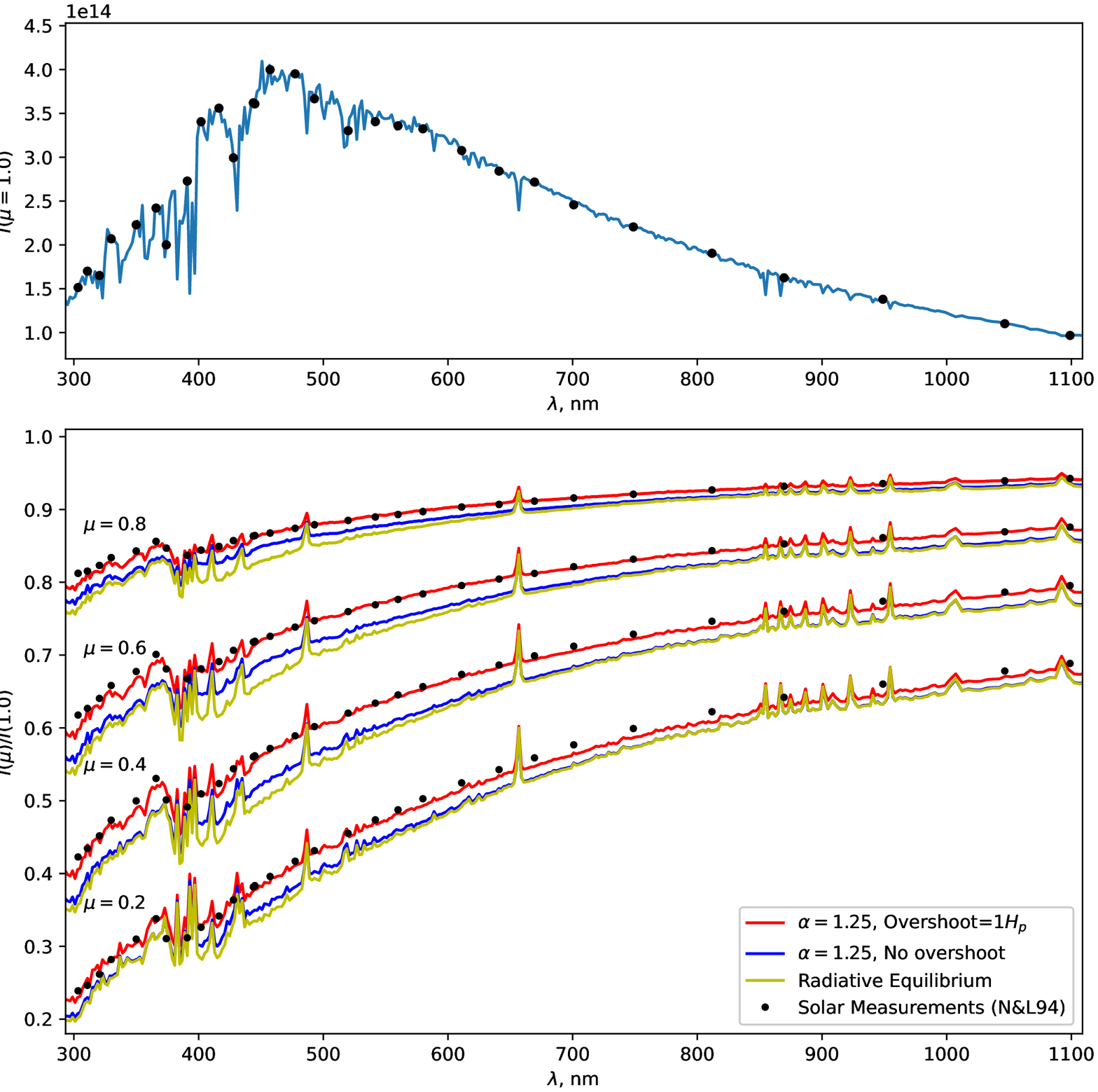}\\
    \caption{Solar spectrum (top panel) and comparison of solar limb-darkening computation with solar measurements by \cite{1994SoPh..153...91N} (bottom panel). Stars in both panels depict the solar measurements, and lines represent computations. Measurements and computations of solar spectrum in the top panel are shown at the disk center. Spectra at selected view angles, namely the $\mu$ = 0.8, 0.6, 0.4, 0.2, normalized to the solar spectrum at the disk center are shown in the bottom panel. Different line types marked in the legend to the figure correspond to the computation with different approximations: RE, with convection ($\alpha =1.25$) and with convection and overshoots. Intensity is measured in $[\rm{erg}~\rm{cm^{-3}}\rm{s^{-1}}\rm{sr^{-1}}]$.}
    \label{fig:solar_observation}
\end{figure*}

In upper panel of Figure~\ref{fig:solar_observation} we show the solar spectrum computed with the MPS-ATLAS code, marking wavelengths at which \cite{1994SoPh..153...91N} measurements were  performed. In the lower panel of Figure~\ref{fig:solar_observation} we show computed and measured values of intensity at several view angles normalized to the disk-center intensity. In addition to our final product (i.e., computations with convection and overshoot) we also plot limb darkening computed assuming RE (i.e., no convection and no overshoot) and limb darkening computed taking convection into account but neglecting overshoot. One can see that MPS-ATLAS solar limb-darkening computations with overshoot accurately agree with the observations at all wavelengths and view angles. 
The convection without overshoot heats the atmosphere in very deep layers, which give only marginal contribution to the emergent radiation. Short-ward of $\sim 400$~nm the continuum opacity increases due to multiple photoionization processes and line haze, so that the contribution from the layers affected by convection can be neglected. Long-ward of $\sim 550$~nm, the temperature sensitivity of the Planck function decreases so that the temperature change due to convection does not noticeably affect limb darkening. Therefore, we only see the deviations between pure RE and convection models between $\sim 400$~nm to $\sim 550$~nm. In comparison to the convection, the overshoot affects the temperature structure of higher layers so that the agreement with measurements significantly deteriorates if overshoot is neglected. 
This result is in line with \cite{2013A&A...554A.118P} who showed that limb darkening computed assuming RE does not agree well with observations. 

\subsection{Computation-to-computation comparison for solar--analog star} \label{ss: validation2}

After testing our computations against the solar measurements, we perform a validation test for a solar-analog star ($M/H=0.0$, $T_{\rm{eff}}=5800~K$, $\log g = 4.5$) against other computations in the Kepler and TESS passbands. 
First, we take the coefficients for the four-parameters limb-darkening law (Eq.~\ref{eq:claret_law}) from the most widely used libraries in the Kepler \citep{2011A&A...529A..75C} and in TESS \citep{2017A&A...600A..30C} passbands for the closest stellar parameters to our solar analog, namely for $M/H=0.0$, $T_{\rm{eff}}=5750~K$, $\log g = 4.5$. While \citet{2011A&A...529A..75C} give also coefficients based on PHOENIX library, we select the coefficients computed based on ATLAS9 library. Using these coefficients, we compute limb darkening curves and refer to them as ``\citetalias{2011A&A...529A..75C}'' (available for Kepler passband) and ``\citetalias{2017A&A...600A..30C}'' (available for TESS passband). 
Second, we take the limb darkening coefficients derived by fitting the same four-parameters law in each passband from our sets and compute limb darkening using these coefficients. We refer to these computations as ``LD1'' and ``LD2'' for the limb darkening computed from set~1 and set~2 libraries, respectively. 

\begin{figure*}[!htb]
    \centering
    \includegraphics[width=0.95\linewidth]{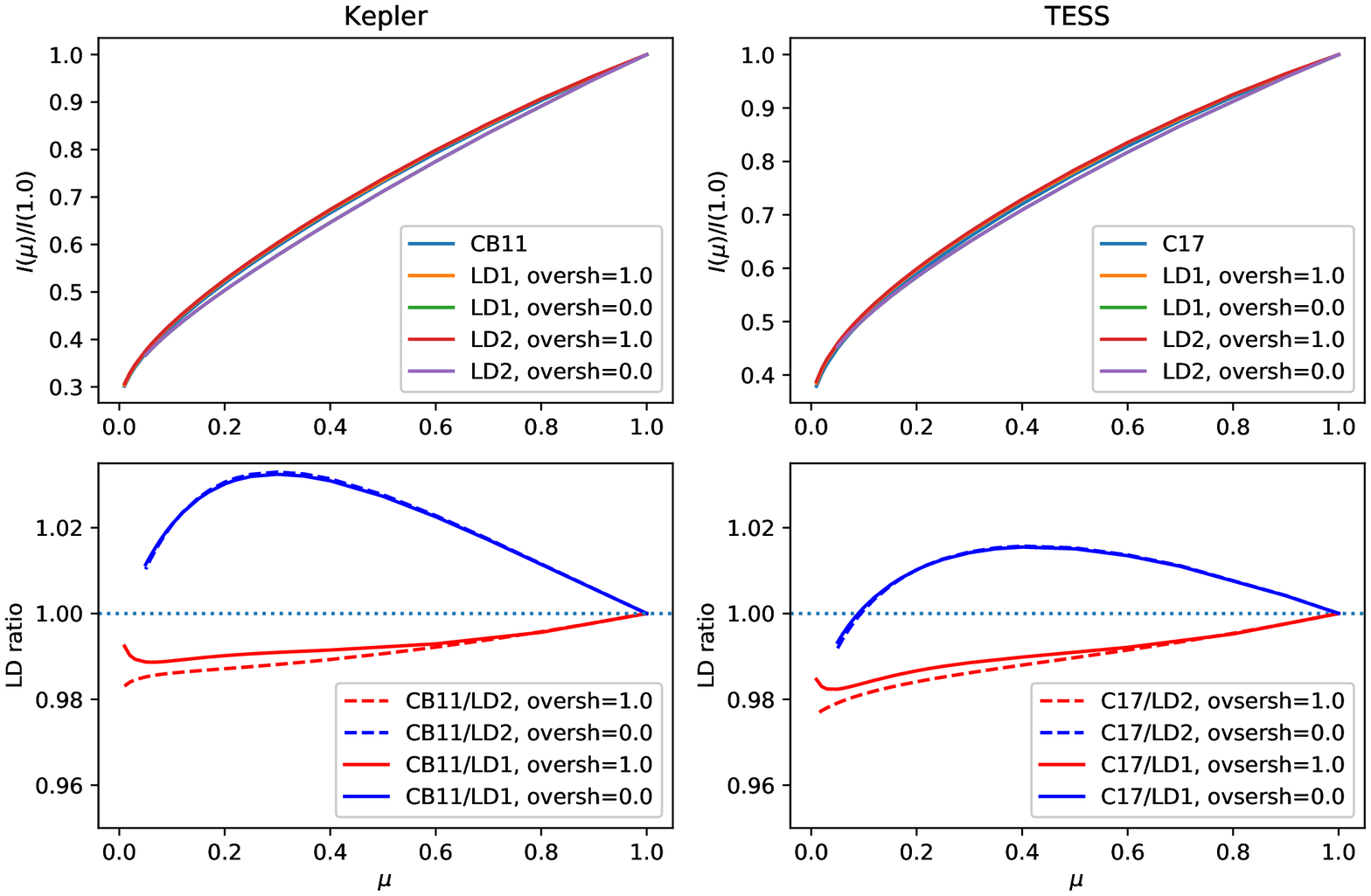}
    \caption{Comparison of the MPS-ATLAS limb-darkening computations to that by \citetalias{2011A&A...529A..75C} (for the Kepler passband, left panels) and by \citetalias{2017A&A...600A..30C} (for the TESS passband, right panels) for a star with solar fundamental parameters. Top panels describe the limb darkening curves, and bottom panels present their ratios. Calculations with MPS-ATLAS (LD1 and LD2) are given, taking the overshoot into account (blue) as well as ignoring it (red). See text for the explanation of the abbreviations. 
    }
    \label{fig:SolarLDComparisonInTwoFilters}
\end{figure*}

In Figure~\ref{fig:SolarLDComparisonInTwoFilters} we compare LD1 and LD2 computed with and without overshoot to \citetalias{2011A&A...529A..75C} and \citetalias{2017A&A...600A..30C} limb  darkening in Kepler and TESS passbands, respectively.
Left panels of Fig.~\ref{fig:SolarLDComparisonInTwoFilters}  show that in the Kepler passband the \citetalias{2011A&A...529A..75C} limb darkening is closer to LD1 and LD2 computed with overshoot. The maximum difference between \citetalias{2011A&A...529A..75C} and our calculations with overshoot is slightly above, $1\%$ close to the stellar limb. At the same time, our computation without overshoot deviates up to $\sim 3\%$ from \citetalias{2011A&A...529A..75C}. Right panels of Fig.~\ref{fig:SolarLDComparisonInTwoFilters} shows that in the TESS passband the maximum difference between our computations and \citetalias{2017A&A...600A..30C} is about $2\%$. Interestingly, the difference between LD1 and LD2, arising mainly due to the difference between \cite{1998SSRv...85..161G} and \citet{2009ARAA..47..481A} abundances, is significantly more prominent for the models with overshoot. 

\begin{figure}[!htb]
    \centering
    \includegraphics[width=0.95\linewidth]{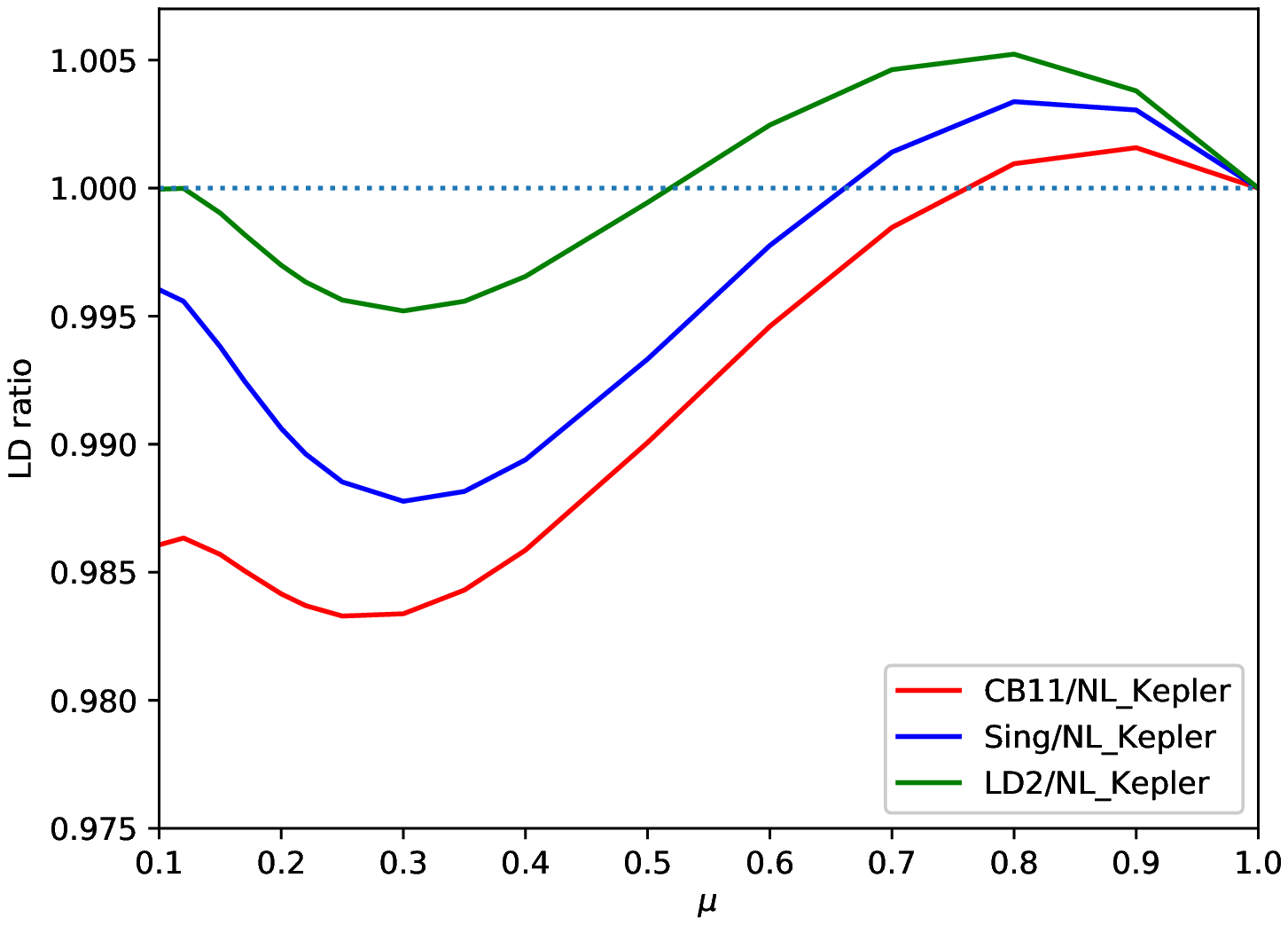}
    \caption{Ratios between limb darkening from LD2, \citetalias{2011A&A...529A..75C}, and \citet{2010A&A...510A..21S} (Sing) to the NL\_Kepler limb darkening. Different colors correspond to different ratios (see legends). }
    \label{fig:SolarLDComparisonInKepler}
\end{figure}

The discrepancy of limb darkening in both passbands between different computations are quite significant and can lead to bias in exoplanet radius determination. In order to understand how accurate our limb darkening (LD1 and LD2) are, we compare them to the data that mimic solar observations in Kepler passband. Since the Sun was never observed in the Kepler passband, we emulate the observations by applying the following trick.
We compute the difference between the solar observations \citep{1994SoPh..153...91N} and our computations with overshoot (from section~\ref{ss: validation1}) at each wavelength between 400~nm and 900~nm where observations are available and for each view angle. The difference monotonously changes with wavelength at each $\mu$. This monotonicity prompts us to interpolate the difference to the MPS-ATLAS wavelength grid in the Kepler passband obtaining corrections to our computations as a function of wavelength and view angle.  After that we apply these corrections to our calculations, convolve them with the Kepler transmission passband and apply the fitting procedure for the  four-parameters limb-darkening law. We refer to the derived limb darkening as ``NL\_Kepler''. In Figure~\ref{fig:SolarLDComparisonInKepler} we present the ratio of LD2 and \citetalias{2011A&A...529A..75C} to the NL\_Kepler. In addition, we present the limb darkening of a solar--analog star in the Kepler filter computed by \cite{2010A&A...510A..21S} using ATLAS9 models. We show that the deviations between our limb darkening and NL\_Kepler reach up to $0.5\%$, while for other previous calculations larger deviations occur especially closer to the limb.  

\subsection{Stellar computation-to-measurements comparison}\label{ss: ld_vs_obs}

Recently, \cite{2018A&A...616A..39M} presented limb darkening coefficients
\begin{equation}
    h_1 = \frac{I_{pb}(0.5)}{I_{pb}(1.0)} = 1 -c(1-2^{-\alpha})
    \label{eq:h1}
\end{equation}
and 
\begin{equation}
    h_2 = \frac{I_{pb}(0.5)}{I_{pb}(1.0)}  - \frac{I_{pb}(0.0)}{I_{pb}(1.0)}  = c 2^{-\alpha}
    \label{eq:h2}
\end{equation}
of the power-2 law derived from light curves for transiting exoplanet systems observed with Kepler. He selected the Kepler targets for which the stellar parameters, such as $M/H$, $T_{\rm{eff}}$, and $\log g$ were determined using high-precision spectroscopy. He compared the obtained coefficients to STAGGER calculations by \citet{2015A&A...573A..90M} interpolated to the fundamental parameters of the Kepler targets \citep[see, Fig.~4 in ][]{2018A&A...616A..39M}.

\begin{figure}[!htb]
    \centering
    \includegraphics[width=0.95\linewidth]{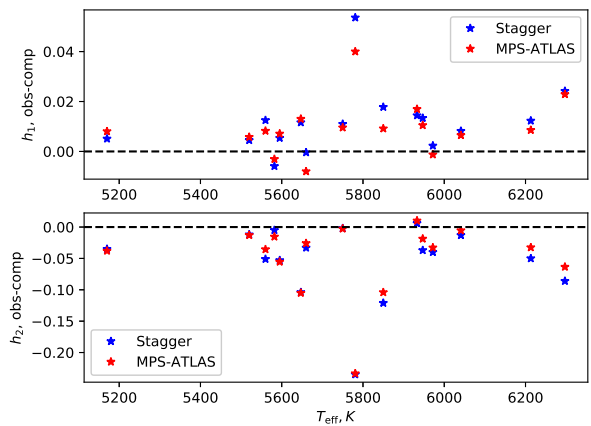}
    \caption{Comparison of the limb darkening coefficients. Star signs show the difference between observations \citep{2018A&A...616A..39M} and computations done with STAGGER (blue) and MPS-ATLAS (red) codes in Kepler passbands. Horizontal dashed line represents on the zero level.}
    \label{fig:h1h2_fromCLV}
\end{figure}

In order to compare our models to observations, we chose to calculate the $h_1$ and $h_2$ coefficients for the same set of stars as in \citet{2018A&A...616A..39M}.
Since 1D MPS-ATLAS computations are very fast, we perform model atmospheres and limb darkening computations directly for the considered Kepler targets, avoiding any interpolation. We directly infer the $h_1$ and $h_2$ coefficients using Eqs.~\ref{eq:h1}--\ref{eq:h2} from our calculations. In Figure~\ref{fig:h1h2_fromCLV} we present the difference between observations and the two computations (STAGGER and MPS-ATLAS) for $h_1$ and $h_2$. For most cases, our computations are closer to observations. However, for several Kepler targets, the 3D STAGGER limb-darkening coefficients show better agreement with observations. We note that, in contrast to the MPS-ATLAS computations, the difference between STAGGER computations and observations also contains uncertainties brought about by the interpolation. Thus, our result does not imply that 1D MPS-ATLAS computations are more accurate than 3D computations, but rather indicates that the combined error of 3D computations and interpolation (which is unavoidable for grids of computationally expensive 3D models) might be even larger than the error of 1D computations.

Interestingly, Fig.~\ref{fig:h1h2_fromCLV} shows that there are systematic differences between the $h_1$ and $h_2$ parameters deduced from observations and from modelling. Namely, $h_1$ values from observations are systematically larger than those from modelling, while $h_2$ values from observations are systematically lower than that from modelling. Note that neither the error bars of measurements \citep[see Table A.1. in][]{2018A&A...616A..39M} nor the computation uncertainties coming from non-ideal stellar parameter determination (private communication with H.-G. Ludwig) can explain the systematic bias of the coefficients. 
In a follow-up paper, we will show that deviations in $h_1$ can be explained by stellar surface magnetic fields (which are neither accounted in MPS-ATLAS nor in STAGGER computations). The interpretation of the deviation in the $h_2$ values is not so clear because the value obtained by fitting the transit light curve of a hot Jupiter may be biased by the extraction procedure (see, Appendix~\ref{aa:h1h2_biases}).
This conclusion is based on using a power-2 transit model to fit simulated transit light curves based on a realistic solar limb-darkening profile. The bias in the $h_1$ values obtain from transit fitting is much lower ($<0.001$) provided that the impact parameter for the transit is $b \loa 0.8$. More work is needed to better understand the best choice of parameters for testing limb-darkening profiles from stellar models against observations of transit light curves for real stars.

\section{Summary and outlook} \label{sect: summary}
In this paper we have employed our recently developed MPS-ATLAS code to compute a new library of stellar model structures and to synthesize stellar limb darkening in different broad passbands (Kepler, TESS, CHEOPS, PLATO) for an extensive and fine grid of stellar parameters. We fit the ``power-2'' and ``4-coefficients'' laws to the computed limb darkening and derived the limb-darkening coefficients. We created two grids of stellar atmospheres and limb darkening parameters with different solar abundances \citep{1998SSRv...85..161G, 2009ARAA..47..481A} and different mixing length parameter considerations (constant or varying for different stellar parameters). We have made the full set of computations (model atmosphere structure, limb darkening, and their coefficients) for our two sets available in CDS. 

Our computations successfully passed different validation tests, such as comparisons with solar measurements, with other computations of stellar limb darkening, and with stellar limb darkening coefficients derived from Kepler observations. We found a very good agreement between our computed solar limb darkening and measurements at different wavelengths and viewing angles. Our computations agree with Kepler measurements at the same level as 3D STAGGER calculations, and for many of the stars even slightly better than STAGGER. However, the sets of coefficients derived from STAGGER and our computations show a systematic offset relative to the observational data. 
This offset can come from ignoring the effect of magnetic activity in limb darkening computations.
In the forthcoming paper, we will quantify the effects of magnetic activity on stellar limb darkening using 3D radiative-magnetohydrodynamics (MHD) simulation with the MURaM code \citep{muram}. Subsequently, we will extend the library presented here to the case of magnetized stars.


\begin{acknowledgement}
We thank the referee, Giuseppe Morello, for constructive and useful comments on the manuscript. 
We are also thankful to H.-G.~Ludwig for our fruitful discussion and to Jesper Schou for useful comments on our manuscript. 
The preliminary PLATO passband transmission curve 
\citep{2019A&A...627A..71M} was kindly made available to us by Martin Pertenais.
N.M.K. and L.G. acknowledge support from the German Aerospace Center (DLR FKZ~50OP1902). V.W. and A.I.S. were funded by the European Research Council (ERC) under the European Union’s Horizon 2020 research and innovation program (grant no. 715947). L.G. acknowledges funding from ERC Synergy Grant WHOLE~SUN~810218 and from NYUAD Institute Grant G1502.
\end{acknowledgement}

\bibliographystyle{aa}
\bibliography{kostogryz}

\clearpage

\begin{appendix} 

\section{High-resolution calculations versus the ODF approach}\label{aa: ODF}
In this section we test the ODF approach, used to produce limb-darkening dependencies presented in our study, against  direct (i.e., without utilizing ODF) calculations on a fine spectral grid (hereafter, high-resolution calculations).

For the high-resolution calculations we computed $I(\lambda, \mu)$ in Eq.~\ref{Eq: filter_convolution} with a spectral resolution of R=500000. For the ODF calculations presented in this study, we used the standard `little' wavelength binning grid \citep{2005MSAIS...8...34C}. In each bin, we first sorted the opacities in ascending order, and then we split the bin further into sub-bins. In each of the sub-bins we then calculated the mean value of the opacity and instead of solving the radiative transfer equation on the high-resolution spectral grid we solved it for each of  the sub-bins. Summing up the weighted values of intensity from each of the sub-bins within the bin (with weights given by the ratios of sub-bin and bin sizes)  we then obtained approximate values of the intensity in the  bins \citep[see the detailed description of the ODF procedure and its performance in][]{2019A&A...627A.157C}. For most of the calculations (see below for the explanation of the exceptions) we utilized the ODF setup proposed by \cite{2003IAUS..210P.A20C} splitting each bin into 12 sub-bins (with 1/10 bin width for the first nine sub-bins with the lowest opacity values, and then with 1/20, 1/30, and 1/60 bin width for the remaining three sub-bins). Such a non-uniform splitting of bins into sub-bins allows an accurate accounting for the cores of strong spectral lines \cite[see][]{2019A&A...627A.157C}. 

\begin{figure}[h!]
    \centering
    \includegraphics[width = 0.99\linewidth]{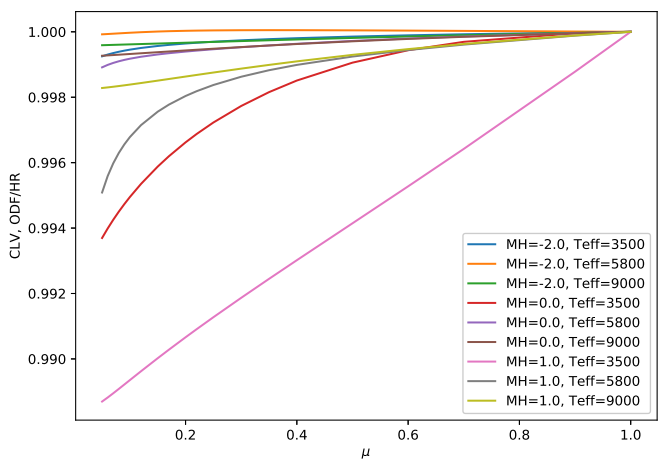}
    \caption{Ratios of intensity values in Kepler passband computed using ODF to that using high resolution opacity (resolving power R=500000). Calculations are performed for several pairs of metallicities and effective temperatures, but the same value of the surface gravity ($\log g = 4.5$).}
    \label{fig:odfHRratio}
\end{figure}

Figure~\ref{fig:odfHRratio} shows that the ODF approach using the setup by \citet{2003IAUS..210P.A20C}  gives  very accurate values of intensity (with errors below 0.2\%) except for cold and/or metal-rich stars. The deterioration of the ODF performance for cold stars is not surprising. The error of the ODF approach has two main components. The first component is due to the replacement of the {\it intensity} averaged over the sub-bin by intensity calculated using {\it opacity} averaged over the sub-bin. The second component is brought about by the failure of the ODF approximation. According to this approximation, the sorting of opacities is the same  over the entire atmospheric region contributing to the emerging intensity \citep[see detailed discussion in][]{1986A&A...167..304E}. The decrease of the sub-bin widths leads to the decrease of the first component of the error (since smaller sub-bins correspond to averaging of the closer values of opacity) but to the increase of the second component (since allocation of frequencies into sub-bins gets more susceptible to the failure of the ODF approximation). For the cold and metal-rich stars, the opacity is brought about by the mixture of lines from different species. These species (in particular, molecules which are the most important opacity source in the atmospheres of cold stars) have different temperature sensitivities and, thus, form in different parts of the stellar atmosphere. This leads to a significant change of the opacity profile within the atmosphere, violating the ODF approximation. Consequently,  the second component of the ODF error dominates over the first one. \citet{2019A&A...627A.157C} showed that the best strategy in such a case is to split the bin into sub-bins of equal width. Here we follow up on this result and in addition to the sub-binning proposed by \cite{2003IAUS..210P.A20C} also utilize an alternative sub-binning, splitting the bins into four sub-bins of equal width (hereafter, optimized sub-binning). 

\begin{figure*}[h!]
    \centering
    \includegraphics[width = 0.48\linewidth]{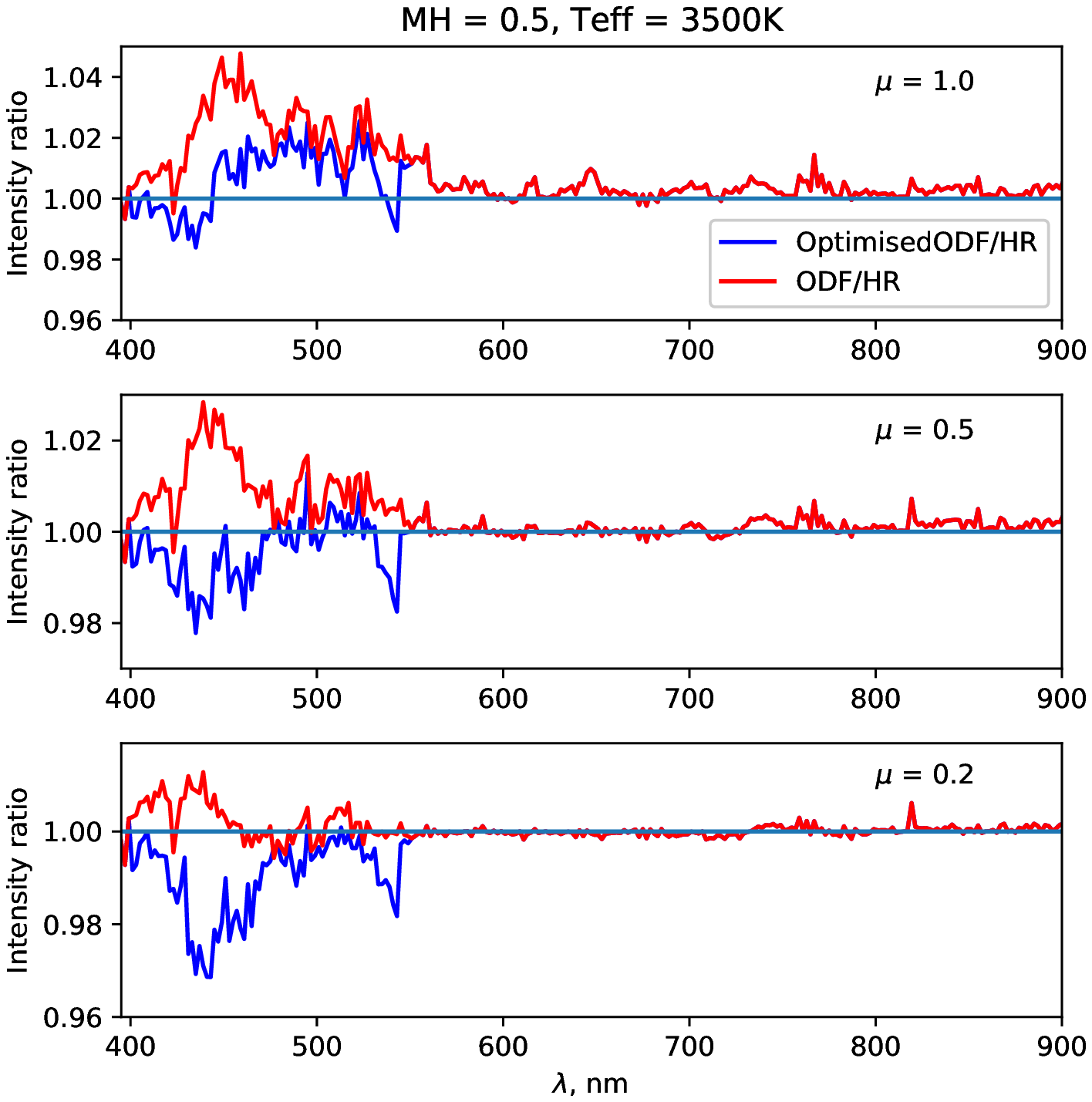}
    \includegraphics[width = 0.48\linewidth]{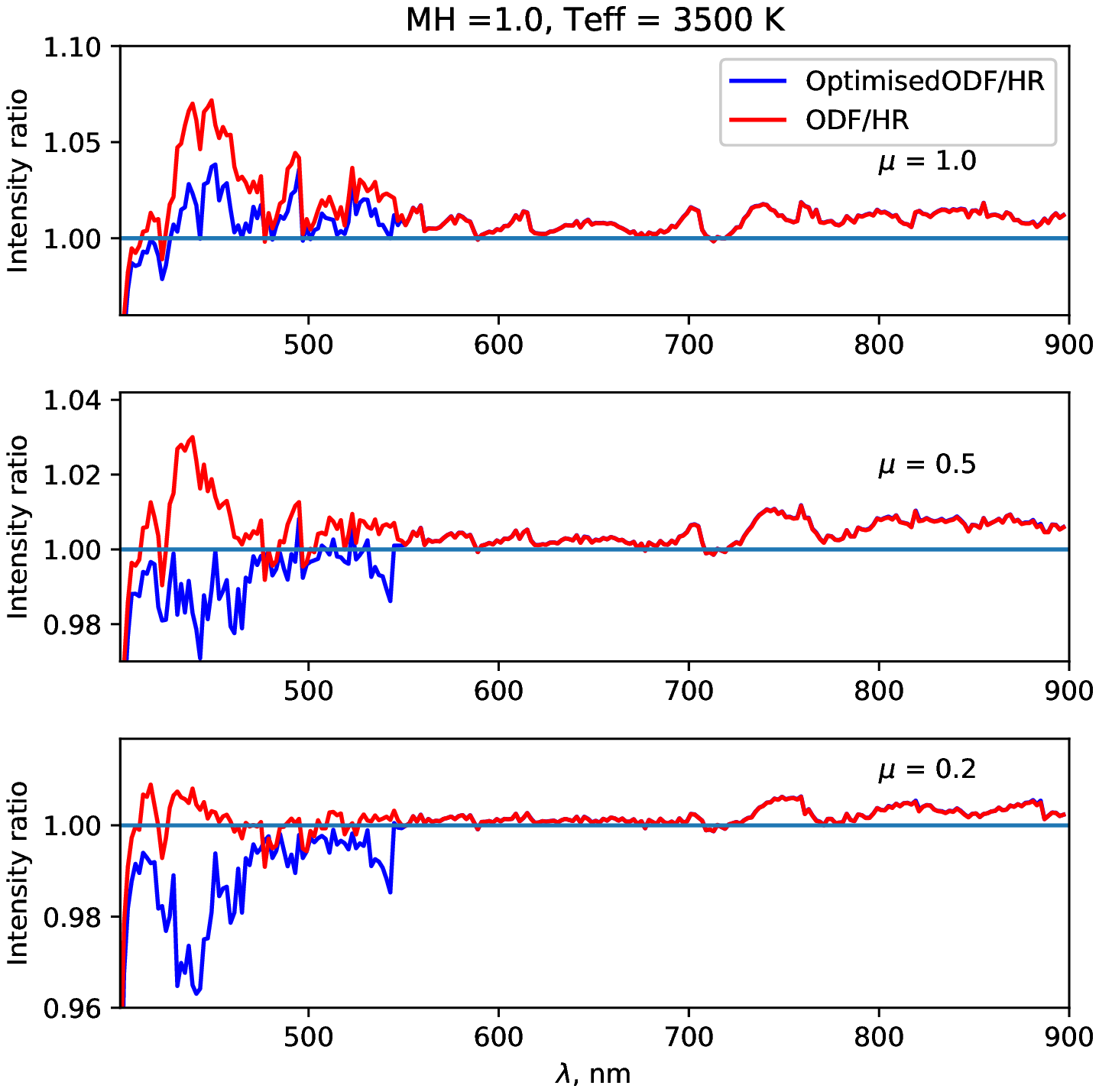}
    \caption{Ratio of emergent intensities at $\mu = 1.0$ (top panels), $\mu=0.5$ (middle panels), and $\mu=0.2$ (bottom panels) calculated using the ODF approach to intensities calculated on a fine spectral grid and then averaged over bins used in the ODF calculations. Shown are the ratios for the model with $M/H=0.5$ (left column) and $M/H=1.0$ (right column), $T_{\rm{eff}}=3500~K$, and $\log g = 4.5$. ODF calculations between $400$ and $550$~nm are performed with two different setups: using sub-binning proposed by \cite{2003IAUS..210P.A20C} (red) and using sub-binning optimized for cold stars (i.e., four sub-bins of equal size, blue). ODF calculations longward $550$~nm are performed only with \cite{2003IAUS..210P.A20C} sub-binning.}
    \label{fig:spectra_ODFvsHRvsOptimizedODF}
\end{figure*}

Figure~\ref{fig:spectra_ODFvsHRvsOptimizedODF} shows a comparison of the performance of the \citet{2003IAUS..210P.A20C} and the optimized sub-binning for stars with $T_{\rm eff}=3500$~K. One can see that the optimized sub-binning performs better for the disk-center calculations (top panels of Fig.~\ref{fig:spectra_ODFvsHRvsOptimizedODF}) but the \citet{2003IAUS..210P.A20C} sub-binning performs better for the near-limb calculations (bottom panels of Fig.~\ref{fig:spectra_ODFvsHRvsOptimizedODF}). This is because photons emitted close to the limb come from a relatively narrow region of stellar atmospheres (in other words, the contribution function for the $\mu=0.2$ intensity is narrower than that for the $\mu=1$ intensity). As a result, the ODF approximation is more precise for the near-limb calculations and smaller sub-bin widths lead to better results by reducing the first component of the ODF error.

\begin{figure}[h!]
    \centering
    \includegraphics[width = 0.99\linewidth]{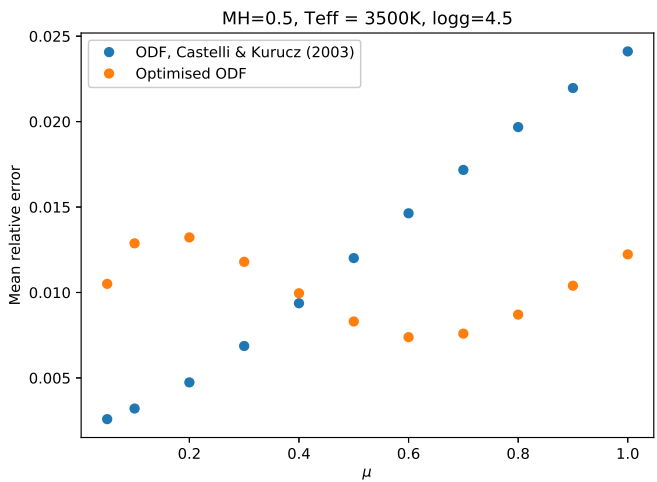}
    \caption{RMS error of the ODF calculations in the $400--550$~nm spectral domain. Shown are errors of calculations with \cite{2003IAUS..210P.A20C} sub-binning (blue) and optimized sub-binning (orange) computed relative to the high-resolution calculations. }
    \label{fig:MeanSquaredError}
\end{figure}

Figure~\ref{fig:MeanSquaredError} shows how the RMS error of ODF calculations (averaged over the $400--550$~nm spectral domain) depends on the view angles. In line with the discussion above, while the error of the ODF calculations with the optimized sub-binning only barely depends on the disk position, the error of the calculations with the \citet{2003IAUS..210P.A20C} sub-binning substantially decreases from the disk center to the limb. 

All in all, we opted to replace calculations based on \citet{2003IAUS..210P.A20C} sub-binning with calculations based on the optimal sub-binning for $\mu \ge 0.5$ for the following values of M/H and $T_{\rm{eff}}$ (independently of $\log g$ values):

\begin{align}
0.3 \le & \rm{M/H} \le 0.45: \,\, T_{\rm{eff}} \le 4200~K; \nonumber \\
0.5 \le & \rm{M/H} \le 0.9: \,\,T_{\rm{eff}} \le 4500~K;  \nonumber \\
1.0 \le & \rm{M/H} \le 1.5: \,\,T_{\rm{eff}} \le 5000~K.
\end{align}

\clearpage
\section{Interpolation errors of stellar limb darkening} \label{aa: interpolation_error}

\begin{figure}[h!]
    \centering
    \includegraphics[width = 0.95\linewidth]{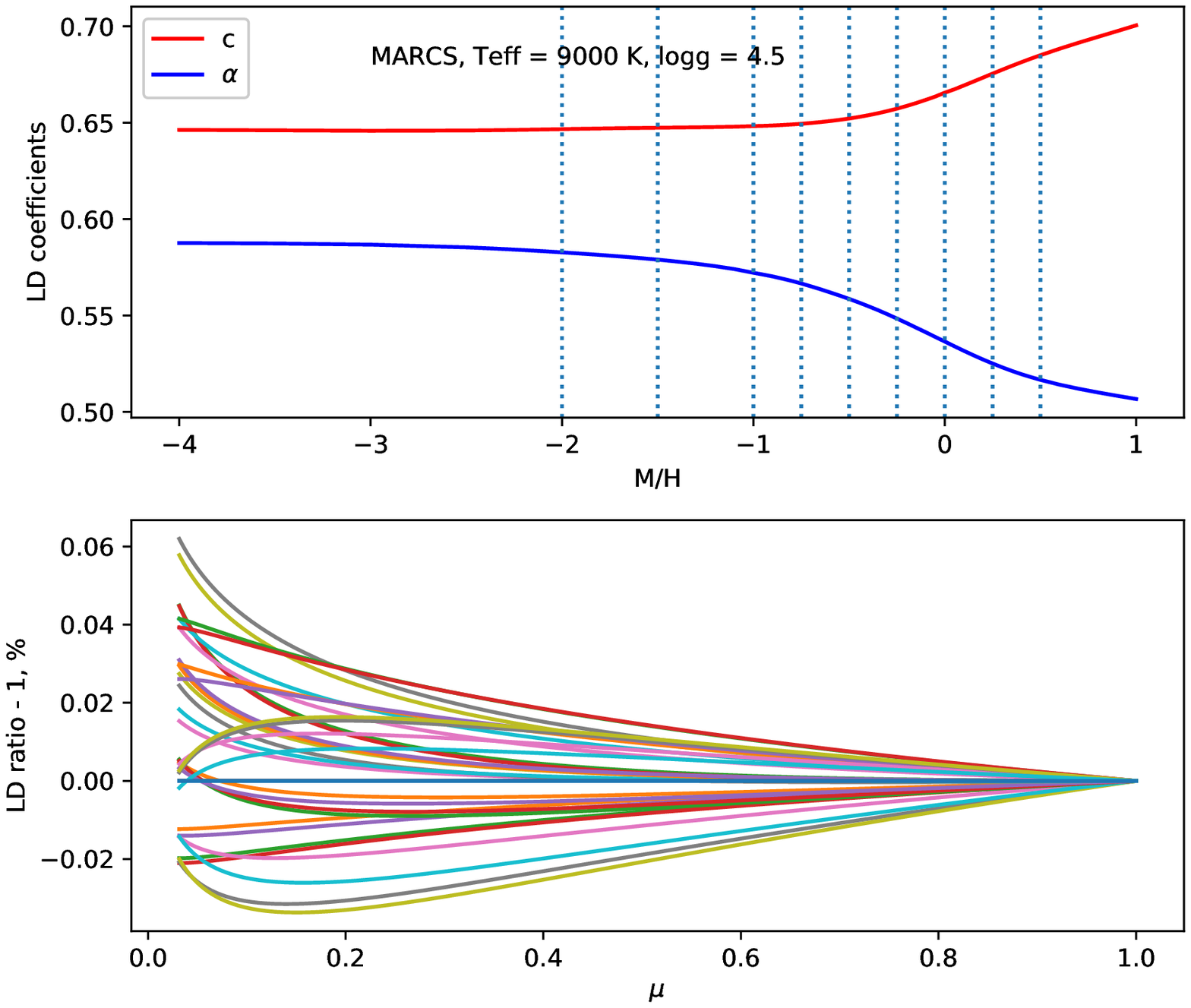}\\
    \includegraphics[width = 0.95\linewidth]{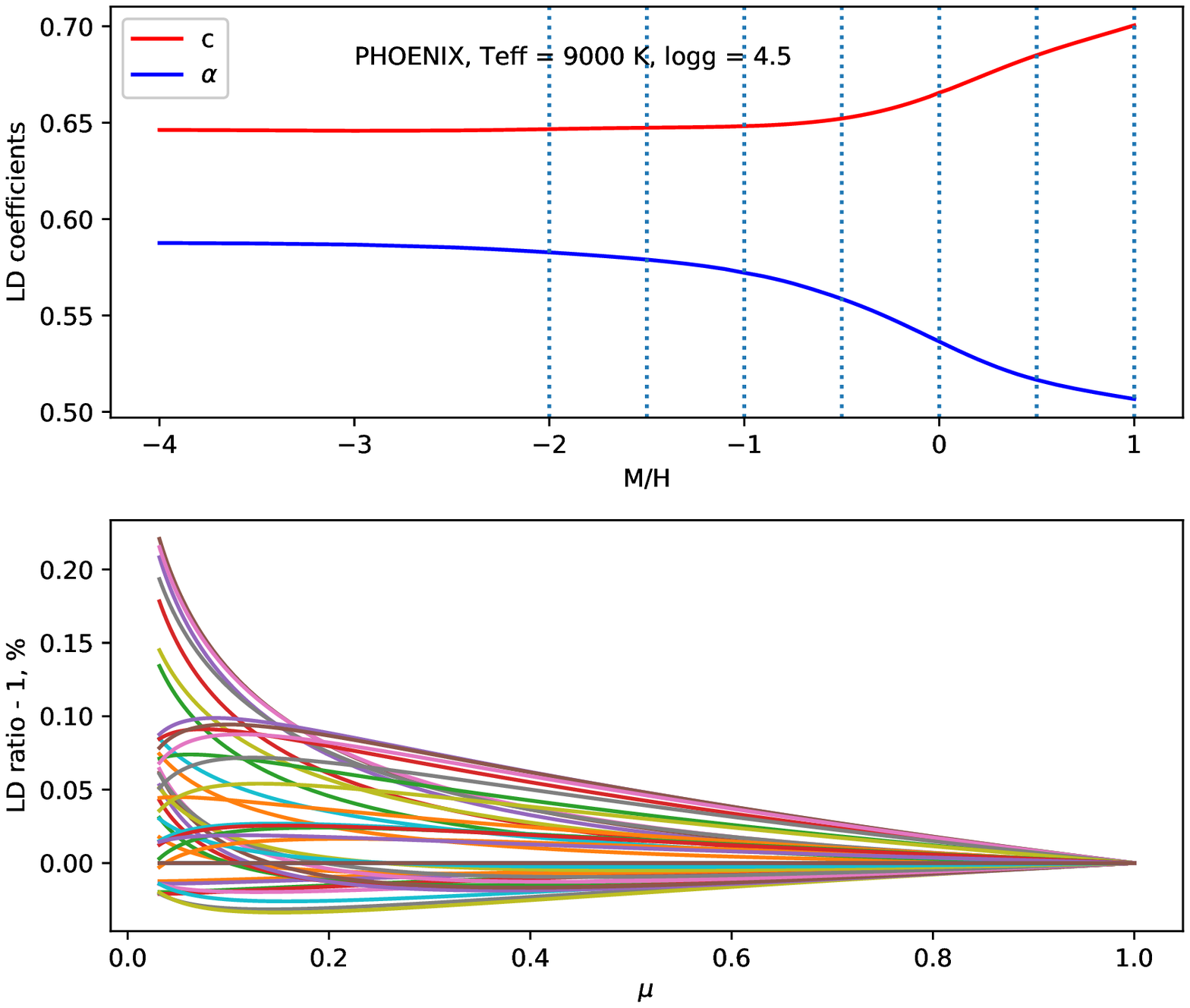}
    \caption{Same as Fig.~\ref{fig:ld_interpolation_error} but for a star with effective temperature of $9000$~K.}
    \label{fig:interpolation_error_9000}
\end{figure}

\begin{figure}[h!]
    \centering
    \includegraphics[width = 0.95\linewidth]{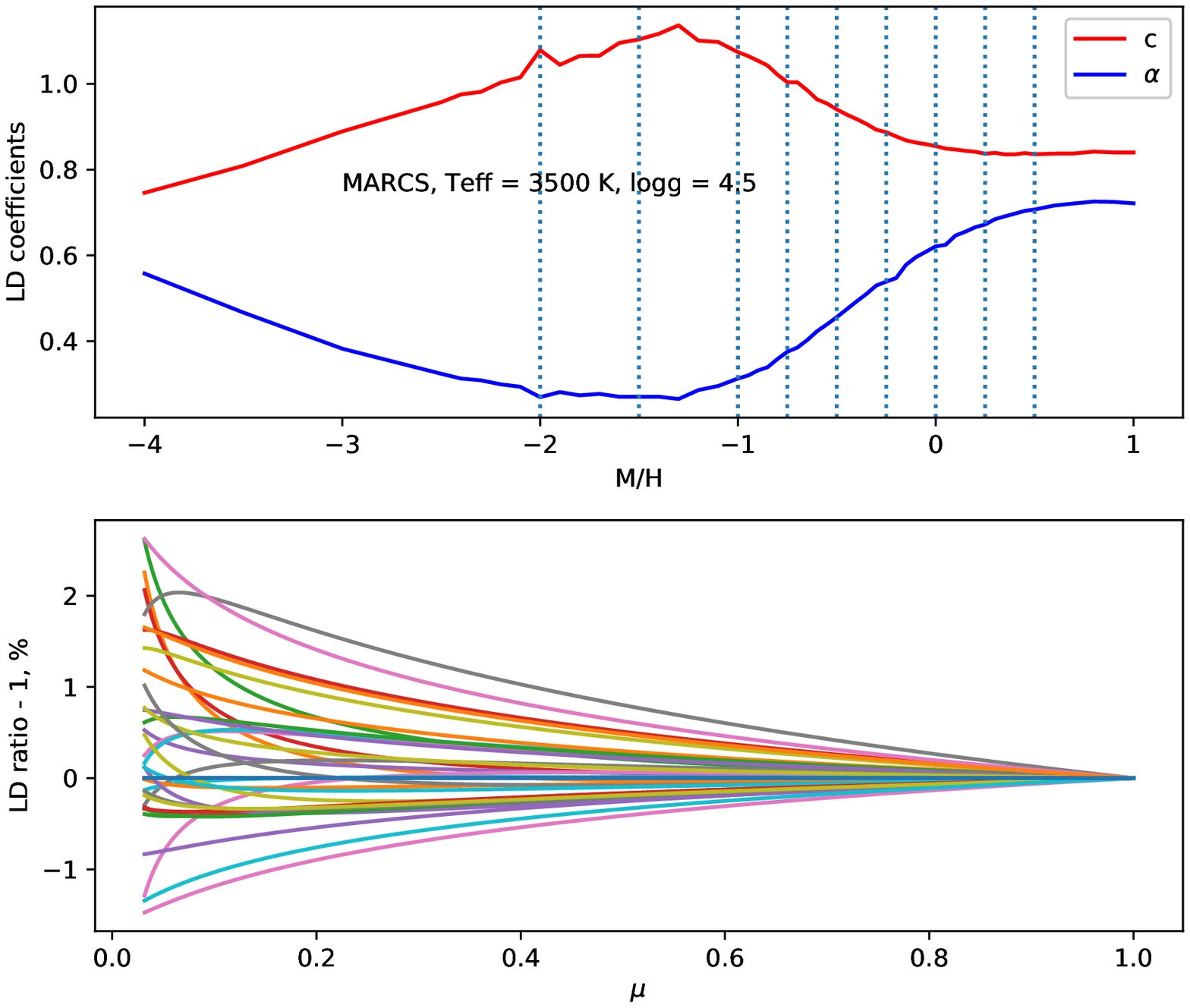}\\
    \includegraphics[width = 0.95\linewidth]{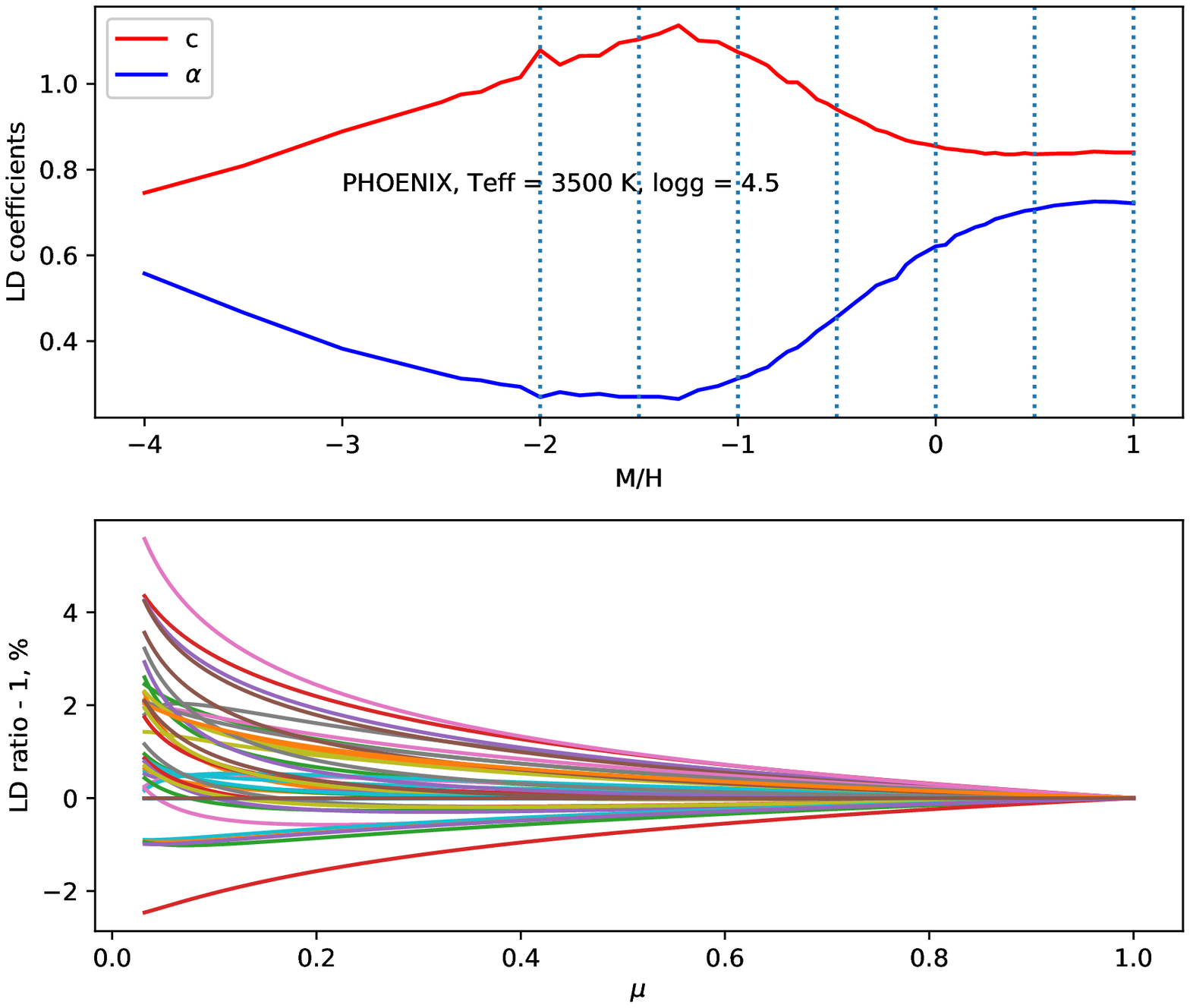}
    \caption{Same as Fig.~\ref{fig:ld_interpolation_error} but for a star with effective temperature of $3500$~K.}
    \label{fig: interpolation_error_3500}
\end{figure}

\begin{figure*}[h!]
    \centering
    \includegraphics[width = 0.49\linewidth]{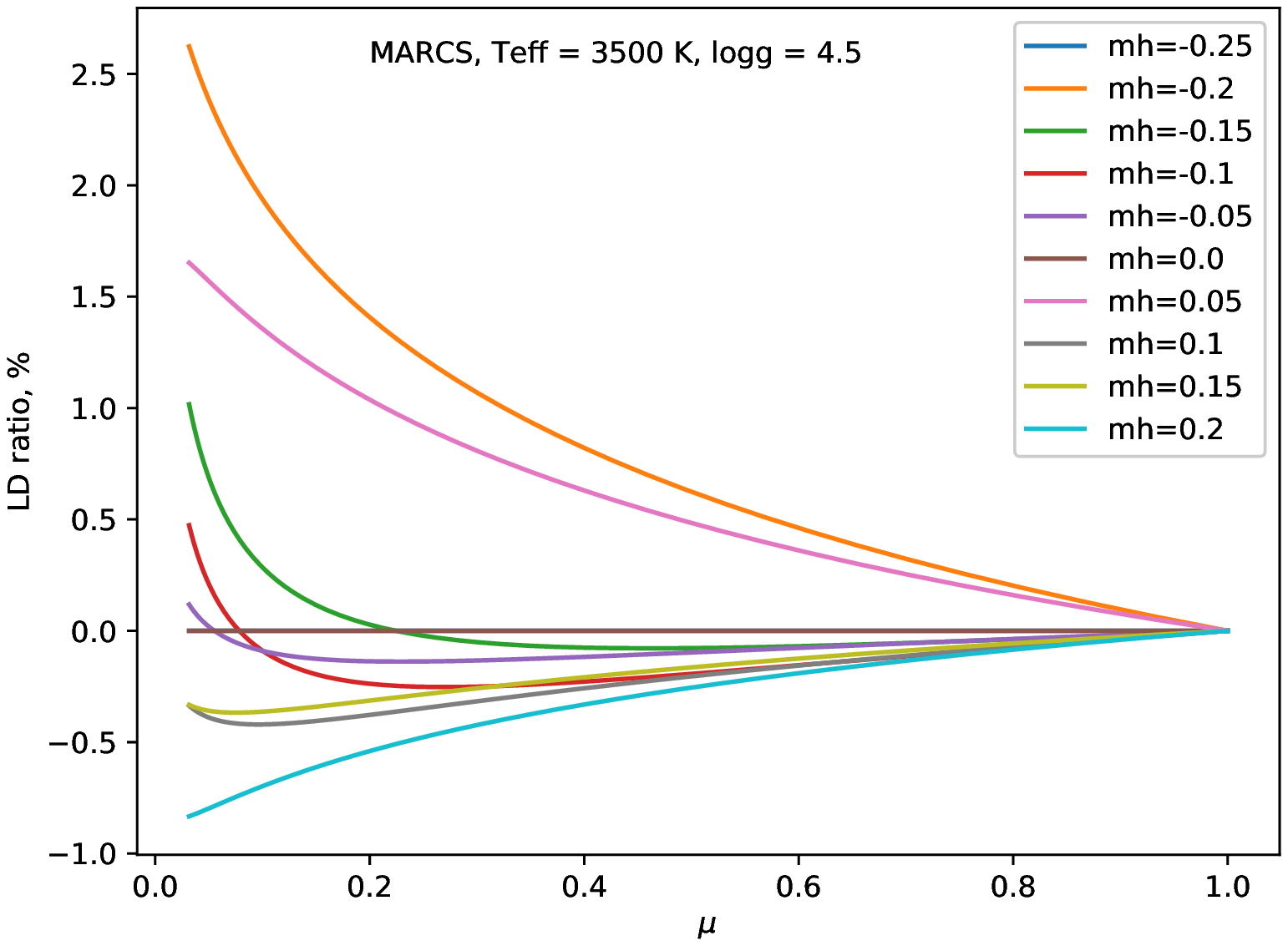}
    \includegraphics[width = 0.49\linewidth]{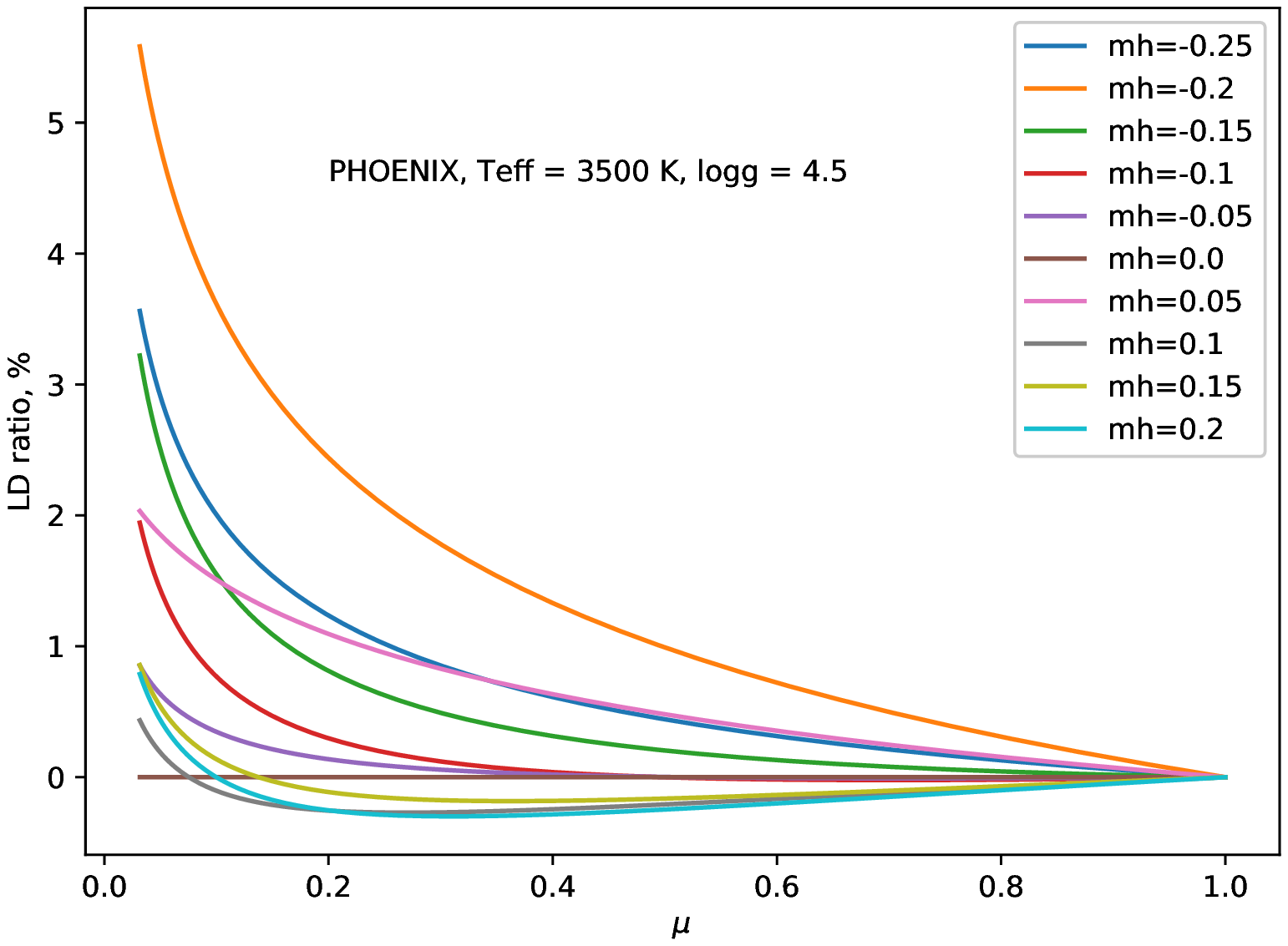}
    \caption{Same as lower panels in Fig.~\ref{fig: interpolation_error_3500} for selected values of metallicity.}
    \label{fig:selected_interpolation_error_3500}
\end{figure*}

In Sect.~\ref{sect:stellar_ld} we discussed the errors in limb darkening curves associated with the interpolation of the power-2 law coefficients from the MARCS and PHOENIX metallicity grids at solar effective temperature (see Fig.~\ref{fig:ld_interpolation_error}). In Figs.\ref{fig:interpolation_error_9000}~and~\ref{fig: interpolation_error_3500} we show corresponding errors for hotter (9000~K) and cooler (3500~K) stars, respectively.
In addition, we present interpolation errors for several selected metallicity values in Fig.~\ref{fig:selected_interpolation_error_3500}.

We note that the $\alpha$ and $c$ are not independent \citep{2018A&A...616A..39M}. As a result, the dependence of the limb darkening coefficients on metallicity contains wiggles (see Fig.~\ref{fig:ld_interpolation_error} in the main text and Fig.~\ref{fig: interpolation_error_3500}) arising from the fitting procedure. This, however, does not imply any abrupt changes in the limb darkening curves since the wiggles in one of the parameters are compensated by the wiggles in another.

\clearpage
\section{Biases of limb darkening coefficients}\label{aa:h1h2_biases}

\begin{figure}[h!]
    \centering
    \includegraphics[width = 0.95\linewidth]{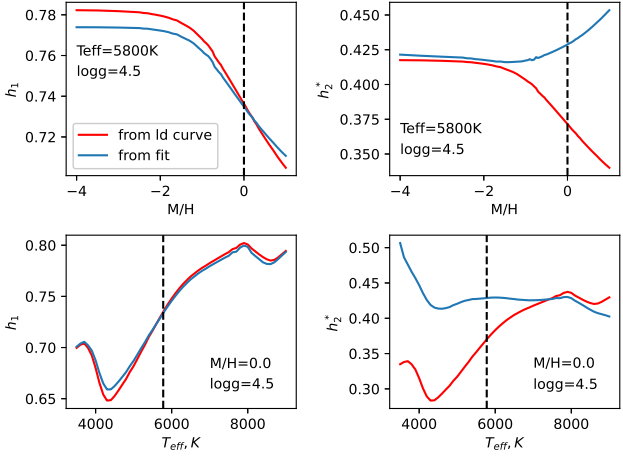}
    \caption{Dependence of $h_1$ (left panels) and $h_2^*$ (right panels, see text for the definition) coefficients on metallicity top panels) and effective temperature (bottom panels). Dashed lines indicate solar value of the metallicity ($\rm{M/H}=0$, top panels) and effective temperature of  5800 K (bottom panels). Shown are  $h_1$ and $h_2^*$ values obtained from the actual intensity profiles (red curves) and power-2 law fit (blue curves).}
    \label{fig:biases_ld_fit}
\end{figure}

The fitting of the limb darkening curve by any parameterized limb darkening law is bound to introduce biases. Fig.~\ref{fig:biases_ld_fit} gives an example of the biases arising from the representation of the limb darkening curve by the power-2 law (see Eq.~\ref{eq: power-2}). The biases are given in terms of the $h_1$ and $h^*_2$ coefficients (see Sect.~\ref{ss: ld_vs_obs}). The $h_1$ coefficient is defined by Eq.~\ref{eq:h1}. Since our limb darkening curves stop at $\mu=0.01$ instead of using the $h_2$ coefficient given by Eq.~\ref{eq:h2} we introduce the $h_2^*$ coefficient:
\begin{equation}
    h_2^* = \frac{I_{pb}(0.5)}{I_{pb}(1.0)}  - \frac{I_{pb}(0.01)}{I_{pb}(1.0)}  = c ( 2^{-\alpha} -   100^{-\alpha}).
    \label{eq:h2_modified}
\end{equation}
We compare the coefficients derived directly from the actual intensity profiles with the coefficients derived from $c$ and $\alpha$ coefficients extracted from power-2 law fit. Figure~\ref{fig:biases_ld_fit} shows that $h_1$ coefficient is stable and can be obtained from the fitting procedure (see left panels of Fig.~\ref{fig:biases_ld_fit}), however, the fitting procedure can not return accurate values of $h_2^*$ (see right panels of Fig.~\ref{fig:biases_ld_fit}).

\onecolumn
\clearpage
\section{Model atmosphere structure} \label{aa: model}

\begin{table}[h!]
\begin{tabular}{ |c|c|c|c|c|c|c| } 
 \hline
 Column Mass & $T_{\rm gas}$ [K] & $P_{\rm gas} [g/cm^3]$ & $n_e, [cm^{-1}]$ & mean $\kappa_{\rm{Ross}}$ & $P_{\rm{rad}} [g/cm^3]$ & $v_{\rm{turb}}$ [cm/s]  \\ 
  \hline
 8.2897116E-04 &  3832.3 & 2.621E+01 & 4.536E+09 & 3.815E-04 & 4.945E-02 & 2.000E+05\\
 1.0291720E-03 &  3846.3 & 3.255E+01 & 5.557E+09 & 4.365E-04 & 5.404E-02 & 2.000E+05\\
 1.2507761E-03 &  3861.5 & 3.955E+01 & 6.675E+09 & 4.935E-04 & 5.708E-02 & 2.000E+05\\
 1.4981936E-03 &  3876.7 & 4.738E+01 & 7.909E+09 & 5.553E-04 & 5.990E-02 & 2.000E+05\\
 1.7754502E-03 &  3892.0 & 5.614E+01 & 9.278E+09 & 6.230E-04 & 6.256E-02 & 2.000E+05\\
 2.0869355E-03 &  3907.5 & 6.599E+01 & 1.080E+10 & 6.974E-04 & 6.506E-02 & 2.000E+05\\
 2.4374832E-03 &  3923.3 & 7.708E+01 & 1.251E+10 & 7.796E-04 & 6.741E-02 & 2.000E+05\\
 2.8324288E-03 &  3939.5 & 8.957E+01 & 1.441E+10 & 8.708E-04 & 6.961E-02 & 2.000E+05\\
 3.2778695E-03 &  3956.4 & 1.037E+02 & 1.655E+10 & 9.713E-04 & 7.151E-02 & 2.000E+05\\
 3.7811817E-03 &  3974.2 & 1.196E+02 & 1.895E+10 & 1.081E-03 & 7.288E-02 & 2.000E+05\\
 4.3511980E-03 &  3992.6 & 1.376E+02 & 2.166E+10 & 1.200E-03 & 7.394E-02 & 2.000E+05\\
 4.9978036E-03 &  4011.4 & 1.580E+02 & 2.472E+10 & 1.332E-03 & 7.476E-02 & 2.000E+05\\
 5.7312986E-03 &  4030.5 & 1.812E+02 & 2.817E+10 & 1.478E-03 & 7.554E-02 & 2.000E+05\\
 6.5628019E-03 &  4049.9 & 2.075E+02 & 3.206E+10 & 1.642E-03 & 7.630E-02 & 2.000E+05\\
 7.5049880E-03 &  4069.6 & 2.373E+02 & 3.645E+10 & 1.825E-03 & 7.706E-02 & 2.000E+05\\
 8.5720521E-03 &  4089.6 & 2.711E+02 & 4.139E+10 & 2.029E-03 & 7.782E-02 & 2.000E+05\\
 9.7798796E-03 &  4109.9 & 3.093E+02 & 4.696E+10 & 2.258E-03 & 7.861E-02 & 2.000E+05\\
 1.1147618E-02 &  4130.6 & 3.525E+02 & 5.324E+10 & 2.508E-03 & 7.901E-02 & 2.000E+05\\
 1.2697192E-02 &  4151.6 & 4.015E+02 & 6.032E+10 & 2.788E-03 & 7.951E-02 & 2.000E+05\\
 1.4451969E-02 &  4172.9 & 4.570E+02 & 6.830E+10 & 3.099E-03 & 8.013E-02 & 2.000E+05\\
 1.6440889E-02 &  4194.4 & 5.199E+02 & 7.731E+10 & 3.439E-03 & 8.051E-02 & 2.000E+05\\
 1.8696647E-02 &  4216.1 & 5.912E+02 & 8.747E+10 & 3.820E-03 & 8.100E-02 & 2.000E+05\\
 2.1252873E-02 &  4237.8 & 6.721E+02 & 9.893E+10 & 4.244E-03 & 8.161E-02 & 2.000E+05\\
 2.4148314E-02 &  4259.7 & 7.636E+02 & 1.118E+11 & 4.719E-03 & 8.237E-02 & 2.000E+05\\
 2.7426245E-02 &  4281.6 & 8.673E+02 & 1.264E+11 & 5.248E-03 & 8.329E-02 & 2.000E+05\\
 3.1135900E-02 &  4303.6 & 9.846E+02 & 1.427E+11 & 5.839E-03 & 8.441E-02 & 2.000E+05\\
 3.5335958E-02 &  4325.8 & 1.117E+03 & 1.612E+11 & 6.489E-03 & 8.536E-02 & 2.000E+05\\
 4.0092969E-02 &  4348.1 & 1.268E+03 & 1.819E+11 & 7.214E-03 & 8.660E-02 & 2.000E+05\\
 4.5478774E-02 &  4370.4 & 1.438E+03 & 2.053E+11 & 8.022E-03 & 8.812E-02 & 2.000E+05\\
 5.1579831E-02 &  4392.7 & 1.631E+03 & 2.317E+11 & 8.911E-03 & 8.963E-02 & 2.000E+05\\
 5.8492898E-02 &  4415.1 & 1.850E+03 & 2.613E+11 & 9.903E-03 & 9.141E-02 & 2.000E+05\\
 6.6323034E-02 &  4437.5 & 2.097E+03 & 2.948E+11 & 1.101E-02 & 9.350E-02 & 2.000E+05\\
 7.5189164E-02 &  4459.9 & 2.378E+03 & 3.324E+11 & 1.224E-02 & 9.590E-02 & 2.000E+05\\
 8.5226074E-02 &  4482.3 & 2.695E+03 & 3.747E+11 & 1.361E-02 & 9.863E-02 & 2.000E+05\\
 9.6585480E-02 &  4504.8 & 3.054E+03 & 4.223E+11 & 1.514E-02 & 1.017E-01 & 2.000E+05\\
 1.0944850E-01 &  4527.5 & 3.461E+03 & 4.760E+11 & 1.683E-02 & 1.050E-01 & 2.000E+05\\
 1.2402293E-01 &  4550.2 & 3.922E+03 & 5.364E+11 & 1.870E-02 & 1.087E-01 & 2.000E+05\\
 1.4053256E-01 &  4572.9 & 4.444E+03 & 6.044E+11 & 2.078E-02 & 1.130E-01 & 2.000E+05\\
 1.5923698E-01 &  4595.7 & 5.036E+03 & 6.810E+11 & 2.309E-02 & 1.176E-01 & 2.000E+05\\
 1.8043159E-01 &  4618.5 & 5.706E+03 & 7.672E+11 & 2.565E-02 & 1.227E-01 & 2.000E+05\\
 2.0444196E-01 &  4641.3 & 6.465E+03 & 8.643E+11 & 2.851E-02 & 1.285E-01 & 2.000E+05\\
 ...& ... &... &... & ...&... &...\\
 6.9362748E+00 &  9506.7 & 2.193E+05  & 4.563E+15 & 5.393E+01 & 5.901E+00 & 2.000E+05 \\
 7.1592382E+00 &  9644.3 & 2.264E+05 & 5.218E+15 & 6.247E+01 & 5.634E+00 & 2.000E+05 \\
 7.4026376E+00 &  9776.4 & 2.341E+05 & 5.925E+15 & 7.177E+01 & 5.138E+00 & 2.000E+05 \\
 \hline
  \end{tabular}\\
 \caption{An example of the model atmosphere structure CDS file}\label{cds: model atmospheres}
\end{table}
  
An example of the model atmosphere structure from set~2 for a star with $M/H = 0.0$, $T_{\rm{eff}}  = 5800$~K, and $\log g = 4.5$ is presented in the Table~\ref{cds: model atmospheres}. All model structures from both set~1 and set~2 are available in CDS.  
For each model we also provide the stellar parameters for which the model was computed, value of mixing-length parameter and overshoot, abundances, and convergence criteria (maximum temperature deviation).

\clearpage
\onecolumn  
\begin{landscape}
\section{Stellar limb darkening} \label{aa:clv}

\begin{table}[h!]
\begin{tabular}{ |c|c|c|c|c|c|c|c|c|c|c|c| } 
 \hline
 \noalign{\smallskip}
 M/H & $T_{\rm{eff}}$ [K] & $\log g$ & Filter & $I_0$ & \multicolumn{7}{c}{$I_\mu/I_0$} \\ 
 &&&&&$\mu=0.9$ &$\mu=0.8$ & $\mu=0.7$ & $\mu=0.6$&$\mu=0.5$ &$\mu=0.4$ & $\mu=0.35$ \\
 \noalign{\smallskip}
 \hline
 \noalign{\smallskip}
 0.0 & 5800 & 3.0 & Kepler& 2.299429e+22 & 0.954169 & 0.90637711 & 0.85627962 & 0.80334817 & 0.74674548 & 0.68510651 & 0.65174239 \\
 0.0 & 5800 & 3.0 & TESS & 3.112463e+22 & 0.96412621 & 0.92624144 & 0.8859303 & 0.84258516 & 0.79528715 &
 0.74259793 & 0.71356034\\
 0.0 & 5800 & 3.0 & CHEOPS & 3.545205e+22 & 0.95291103 & 0.90391565 & 0.85269494 & 0.79875431 & 0.74130115 &
 0.6790307 & 0.64545813\\
0.0 & 5800 & 3.0 & PLATO & 3.130497e+22 & 0.95469595 & 0.90760392 & 0.85841208 & 0.8066422 & 0.75153502 &
 0.69184899 & 0.65968978\\
0.0 & 5800 & 3.5 & Kepler & 2.289401e+22 & 0.95424847 & 0.90635922 & 0.85594453 & 0.8024339 & 0.74496612 &
 0.68222192 & 0.64827821\\
0.0 & 5800 & 3.5 & TESS & 3.103043e+22 & 0.96382873 & 0.92551012 & 0.88460208 & 0.84047679 & 0.79221866 &
 0.73847001 & 0.70893979\\
0.0 & 5800 & 3.5 & CHEOPS & 3.528353e+22 & 0.95308098 & 0.90407529 & 0.85261873 & 0.7981719 & 0.73991464 &
 0.67658253 & 0.64244276\\
0.0 & 5800 & 3.5 & PLATO & 3.127982e+22 & 0.95456711 & 0.90719646 & 0.85754018 & 0.80508501 & 0.74905415 &
 0.68825172 & 0.65551985\\
\noalign{\smallskip}
 \hline
  \end{tabular}\\

Continuation of the table for other $\mu$\\

\begin{tabular}{|c|c|c|c|c|c|c|c|c|c|c|c|} 
\hline
   M/H & $T_{\rm{eff}}$ [K] & $\log g$ & Filter &  $\mu=0.3$ & $\mu=0.25$ & $\mu=0.22$& $\mu=0.2$ & $\mu=0.17$ & $\mu=0.15$ & $\mu=0.12$ & $\mu=0.1$ \\ 
     \hline
  0.0 & 5800 & 3.0 & Kepler& 0.61618003 & 0.57792628 & 0.55344507 & 0.53640692 & 0.50968056 & 0.49103299 & 0.46172275 & 0.44119012\\
  0.0 & 5800 & 3.0 & TESS&  0.68222389 & 0.64809847 & 0.62604762 & 0.61060798 & 0.58623611 & 0.56911445 & 0.54197103 & 0.52273501\\
  0.0 & 5800 & 3.0 & CHEOPS&  0.60977649 & 0.5715094 & 0.54707913 & 0.53010233 & 0.5035126 & 0.48498787 & 0.45591319 & 0.43557476\\
  0.0 & 5800 & 3.0 & PLATO &  0.6255252 & 0.58889661 & 0.56551179 & 0.54925656 & 0.52377818 & 0.50600397 & 0.47803951 & 0.45840362\\
  0.0 & 5800 & 3.5 & Kepler &  0.61219941 & 0.57360478 & 0.54907004 & 0.53208474 & 0.50559934 & 0.48722974 & 0.45849677 & 0.43841199\\
  0.0 & 5800 & 3.5 & TESS &  0.67721969 & 0.64292727 & 0.62093974 & 0.60563144 & 0.58161047 & 0.56482857 & 0.53832829 & 0.51956651\\
  0.0 & 5800 & 3.5 & CHEOPS &  0.60624819 & 0.5676317 & 0.54313532 & 0.52619924 & 0.49982598 & 0.48155844 & 0.45302328 & 0.4331036\\
  0.0 & 5800 & 3.5 & PLATO &  0.62084566 & 0.58387002 & 0.56041369 & 0.54419016 & 0.51890371 & 0.5013622 & 0.47389123 & 0.45464392\\
     \hline
\end{tabular}\\
 
Continuation of the table for other $\mu$\\

\begin{tabular}{|c|c|c|c|c|c|c|c|c|c|c|c|} 
\hline
 M/H & $T_{\rm{eff}}$ [K] & $\log g$ & Filter & $\mu=0.09$ & $\mu=0.08$ & $\mu=0.07$ & $\mu=0.06$ & $\mu=0.05$ & $\mu=0.03$ & $\mu=0.02$ & $\mu=0.01$  \\
\hline
0.0 & 5800 &3.0 &Kepler& 0.43056741 & 0.41964925 &0.40835594 &0.39655563 &0.38402882 &0.35498364& 0.33629603& 0.30979966\\
0.0 &5800 &3.0 &TESS& 0.51268769 &0.50227727& 0.4914053&  0.47991661& 0.46756168& 0.43822948& 0.41879704 &0.39029907\\
0.0 &5800 &3.0 &CHEOPS& 0.42506164 &0.41426255 &0.40309925 &0.39144218 &0.3790755 &0.35043099 & 0.33202249& 0.30596518\\
0.0& 5800 &3.0& PLATO& 0.4482219& 0.43773728& 0.42686981& 0.41549016 &0.40338606 &0.3752644
& 0.35716206& 0.33152686\\
0.0 &5800& 3.5& Kepler& 0.4280108&  0.4172993 & 0.40618727 &0.39453405 &0.38211566 &0.35317981&
0.3344814 & 0.30779138\\
0.0 &5800 &3.5 &TESS &0.50975197& 0.49956206 &0.48889256 &0.47758607 &0.46539707& 0.43640485&
 0.41719137 &0.38898781\\
0.0 &5800 &3.5 &CHEOPS& 0.42279639 &0.41218763& 0.40118839& 0.38965983 &0.37738074 &0.34878915&
 0.33032474& 0.30399372\\
0.0 &5800 & 3.5& PLATO &0.44465646 &0.43435494 &0.42365052& 0.41240625& 0.40040552& 0.37239446&
 0.35428029 &0.32845515\\
\hline
\end{tabular}
\caption{Example of CDS table for stellar limb darkening from set~2 at 24 view angles.}\label{cds:ld}
\end{table}
\end{landscape}

\section{Parameterized stellar limb darkening}
\label{aa: LD_coefficients}
\begin{table}[h!]
\begin{tabular}{|c|c|c|c|c|c|c|c|c|c|} 
\hline
 M/H & $T_{\rm{eff}}$ [K] & $\log g$ & Filter & $\alpha$ & $c$ & $a_1$ & $a_2$ & $a_3$ & $a_4$  \\
\hline
0.0 & 5800 & 3.0 & Kepler & 0.73117934 & 0.62140873 & 0.39036346 & 0.61010164 & -0.34607454 & 0.07746202\\
0.0 & 5800 & 3.0 & TESS & 0.66372177 & 0.54465094 & 0.49676392 & 0.29724876 & -0.12308084 & -0.01209308 \\
0.0 & 5800 & 3.0 & CHEOPS & 0.73149898 & 0.63597503 & 0.38652769 & 0.58555249 & -0.29821916 & 0.06131728\\
0.0 & 5800 & 3.0 & PLATO & 0.70441783  & 0.63208348 & 0.40624144 & 0.48465071 & -0.21847478 & 0.03848722\\
0.0 & 5800 & 3.5 & Kepler & 0.73088466 & 0.62965908 & 0.43258289 & 0.40764303 & -0.03294042 & -0.07099229\\
0.0 & 5800 & 3.5 & TESS & 0.66039557 & 0.55885557 & 0.52725347 & 0.12328191 & 0.15113149 & -0.13983767 \\
0.0 & 5800 & 3.5 & CHEOPS & 0.73160528 & 0.64279276 & 0.42982258 & 0.38472421 & 0.01052298 & -0.08538399\\
0.0 & 5800 & 3.5 & PLATO & 0.70519720 & 0.64126459 & 0.44370735 & 0.30389336 & 0.06255640 & -0.09404369 \\
\hline
\end{tabular}
\caption{Example of CDS table for stellar limb darkening coefficients from set~2.}\label{cds:ld_coeff}
\end{table}

\end{appendix}

\end{document}